\documentclass[pdflatex,sn-mathphys-num]{sn-jnl}

\usepackage{graphicx}%
\usepackage{multirow}%
\usepackage{amsmath,amssymb,amsfonts}%
\usepackage{amsthm}%
\usepackage{mathrsfs}%
\usepackage[title]{appendix}%
\usepackage{xcolor}%
\usepackage{textcomp}%
\usepackage{manyfoot}%
\usepackage{booktabs}%
\usepackage{algorithm}%
\usepackage{algorithmicx}%
\usepackage{algpseudocode}%
\usepackage{listings}%
\usepackage{mathtools}
\usepackage{placeins}

\theoremstyle{thmstyleone}%
\newtheorem{theorem}{Theorem}
\newtheorem{prop}{Proposition}[section]
\theoremstyle{thmstyletwo}%

\theoremstyle{thmstylethree}%

\newcommand{\manif}{\mathcal{M}}
\newcommand{\xbf}{\mathbf{x}}
\newcommand{\Abf}{\mathbf{A}}

\newcommand{\hbf}{\mathbf{h}}
\newcommand{\ybf}{\mathbf{y}}
\newcommand{\Ybf}{\mathbf{Y}}
\newcommand{\sbf}{\mathbf{s}}
\newcommand{\cbf}{\mathbf{c}}
\newcommand{\Mbf}{\mathbf{M}}
\newcommand{\Fbf}{\mathbf{F}}
\newcommand{\Sbf}{\mathbf{S}}
\newcommand{\phibf}{\boldsymbol{\phi}_0}

\newcommand{\phibfn}{\phibf^{\textnormal{obs}}}
\newcommand{\sigbf}{\boldsymbol{\Sigma}}

\newcommand{\Zbf}{\mathbf{Z}}
\newcommand{\qbf}{\mathbf{Q}}
\newcommand{\vbf}{\mathbf{v}}
\newcommand{\Vbf}{\mathbf{V}}
\newcommand{\Kbf}{\mathbf{K}}
\newcommand{\Pbf}{\mathbf{P}}
\newcommand{\ombf}{\boldsymbol{\omega}}
\newcommand{\Ibf}{\mathbf{I}}

\newcommand{\ubf}{\mathbf{u}}
\newcommand{\Lbf}{\mathbf{L}}

\newcommand{\param}{\boldsymbol{\beta}}
\newcommand{\Vs}{V}
\newcommand{\lm}{L^2(\manif)}
\newcommand{\hm}{H^2(\manif)}
\newcommand{\mug}{\mu_g}
\newcommand{\eqdef}{\vcentcolon=}
\newcommand{\bff}{\mathbf{b}}
\newcommand{\kbf}{\mathbf{k}}
\newcommand{\tri}{\mathrm{T}}

\newcommand{\diag}{\text{Diag}}
\newcommand{\alp}{\alpha}
\newcommand{\Scal}{\mathcal{S}}
\newcommand{\Gbf}{\mathbf{G}}
\newcommand{\Rbf}{\mathbf{R}}
\newcommand{\rbf}{\mathbf{r}}
\newcommand{\hh}{\mathbf{h}}
\newcommand{\Dbf}{\mathbf{D}}
\newcommand{\ang}{\delta}
\newcommand{\llik}{\mathcal{L}}
\newcommand{\amalp}{a_{\noise, \alp}^{(m)}}
\newcommand{\amalpzero}{a_{0, \alp}^{(m)}}
\newcommand{\amzero}{a_{0}^{(m)}}
\newcommand{\amnoise}{a_{\noise}^{(m)}}
\newcommand{\Amzero}{A}

\newcommand{\noise}{\tau}
\newcommand{\eps}{E}
\newcommand{\Ebf}{\mathbf{E}}
\newcommand{\Bbf}{\mathbf{B}}
\newcommand{\bbf}{\mathbf{b}}
\newcommand{\Nbf}{\mathbf{N}}
\newcommand{\azero}{a_0}
\newcommand{\anoise}{a_\noise}\newcommand{\wbf}{\mathbf{w}}
\newcommand{\unoisealp}{\ubf_{\noise, \alp}}
\newcommand{\uzeralp}{\ubf_{0, \alp}}

\begin{document}

\title{Spline Interpolation on Compact Riemannian Manifolds}


\author*[1]{\fnm{Charlie} \sur{Sire}}\email{charlie.sire@minesparis.psl.eu}

\author[1]{\fnm{Mike} \sur{Pereira}}\email{mike.pereira@minesparis.psl.eu}

\author[1]{\fnm{Thomas} \sur{Romary}}\lastemail{thomas.romary@minesparis.psl.eu}

\affil*[1]{\orgdiv{Centre for geosciences and geoengineering}, \orgname{Mines Paris, PSL University}, \city{Fontainebleau}, \postcode{77300},  \country{France}}


\abstract{Spline interpolation is a widely used class of methods for solving interpolation problems by constructing smooth interpolants that minimize a regularized energy functional involving the Laplacian operator. While many existing approaches focus on Euclidean domains or the sphere, relying on the spectral properties of the Laplacian, this work introduces a method for spline interpolation on general manifolds by exploiting its equivalence with kriging. Specifically, the proposed approach uses finite element approximations of random fields defined over the manifold, based on Gaussian Markov Random Fields and a discretization of the Laplace-Beltrami operator on a triangulated mesh. This framework enables the modeling of spatial fields with smooth variations and local anisotropies via domain deformation. The method is first validated on the sphere using both analytical test cases and a pollution-related study, and is compared to the classical spherical harmonics-based method. Additional experiments on the surface of a cylinder further illustrate the generality of the approach.}


\keywords{Spline interpolation, Riemannian manifold, Gaussian Markov Random Fields, Anisotropy, Finite elements}

\maketitle

	\section{Introduction}
\label{intro}

In the context of spatially distributed datasets, interpolation methods are especially relevant, as they aim to construct a function defined over the entire domain that exactly matches the values at the observation points. More precisely, let us consider $n$ observations of a phenonemon, denoted by $\left(y_i\right)_{i=1}^n \in \mathbb{R}^n$, measured at points $\Scal = \left(\sbf_i\right)_{i=1}^n$ in a spatial domain $\mathcal{X}.$ The objective is to find a function $ u : \mathcal{X} \rightarrow \mathbb{R} $, such that $u(\sbf_i) = y_i,~1\leq i \leq n.$  

Kriging is a widely used interpolation method for this type of problem. Originating from geostatistics, it is based on the Best Linear Unbiased Predictor (BLUP) framework, providing not only point predictions but also estimates of uncertainty through a covariance structure that encodes spatial dependence. Spline interpolation represents another widely used class of methods for tackling interpolation problems. Among its most general formulations is the thin-plate splines (TPS,~\cite{wahba1990spline}), based on variational principles. This method constructs a smooth interpolant by minimizing a regularized energy functional that incorporates the Laplacian operator $-\Delta$ of the function, thereby penalizing excessive roughness. The approach has a compelling physical interpretation, as it models the bending behavior of a thin elastic plate fixed at specific data points. It involves solving a minimization problem under the interpolation constraints. The TPS (referred to simply as splines in the following) will be the main focus of this manuscript, as their associated minimization problem can be easily formulated within various types of spaces $\mathcal{X}$. Moreover, this spline interpolation has been shown to be equivalent to kriging when using a particular covariance kernel and prior trend function that are related to the eigenvalues and eigenvectors of the Laplacian operator $-\Delta$. This equivalence was first established in the pioneering work of Matheron~\cite{matheron}, and later applied by Olivier Dubrule for the cases $\mathcal{X} = \mathbb{R}$ and $\mathcal{X} = \mathbb{R}^2$~\cite{DUBRULE1984327, dubrule1983two}, as well as in several subsequent studies in $\mathbb{R}^d$~\cite{myers1988interpolation, myers1992kriging}.

However, many real-world datasets are naturally distributed over non-Euclidean domains, such as surfaces or manifolds, with numerous examples found in environmental sciences~\cite{Menafoglio03072021, diaz1999nonlinear, menafoglio2022mathematical} and geosciences~\cite{menafoglio2015object}. One of the most representative cases of surface-based data observation involves measurements taken on the Earth's surface, which is inherently spherical. In such cases, relying on Euclidean distance to model correlations between points is no longer appropriate, since standard covariance models may become ill-defined. As a result, extensive research has compared spline interpolation and kriging on the sphere, beginning with the foundational work of Wahba~\cite{wahba1981spline}, and subsequently expanded by numerous studies in practical situations~\cite{keller2019thin, dunitz2023thin, bonabifard2022semi}. But these works specifically exploit the geometric properties of the
sphere, by expanding the random fields into bases of deterministic functions known as spherical harmonics, and then do not straightforwardly extend to general, arbitrary surfaces. The theory of Reproducing Kernel Hilbert Spaces (RKHS) is then needed to obtain the equivalence between splines and kriging interpolation to general Riemannian manifolds, as formally established in~\cite{kim2001splines}. The focus of this manuscript is therefore the study of Gaussian random fields defined on general compact Riemannian manifolds, equipped with suitable covariance functions, to develop a spline interpolation framework for data distributed over surfaces.

The framework introduced by Lindgren~\cite{Lindgren} and later extended by Pereira~\cite{pereira:tel-02499376} forms the foundation of our study, focusing on Gaussian random fields defined on Riemannian manifolds, denoted here by $\mathcal{X} = \manif$.  His approach relies on finite element approximations of random fields over $\manif$, introducing Gaussian Markov Random Fields (GMRFs, \cite{rue2005gaussian}), that are defined on a mesh of the manifold $\manif$. In our context, this framework enables the discretization of the Laplace-Beltrami operator on the triangulated domain, leading to a finite set of identifiable eigenvalues~\cite{lang2023galerkin}, as will be detailed in this paper.

A possible application of our work is to support the modeling of functions with local anisotropies, as presented in~\cite{pereira:tel-02499376}, through the deformation of the spatial domain. The key idea is that a non-stationary field defined on a domain can be rendered stationary by applying a suitable deformation~\cite{sampson1992nonparametric}. Identifying an appropriate deformation thus allows the problem to be recast in a stationary framework on the transformed domain. Then, although classical spline interpolation inherently assumes stationarity, domain deformation provides a way to construct splines that account for local anisotropies and work with non-stationary covariance functions.

The contribution of this work is to propose a method for solving the spline interpolation problem on compact surfaces beyond the spherical case, extending it to connected orientable Riemannian manifolds. This framework is specifically dedicated to predicting phenomena that vary smoothly on the surface. Within the Riemannian environment, splines with local anisotropies can be effectively modeled, and their finite-element approximation results in faster computation times compared to classical spline predictions.

The outline of this paper is as follows. Section~\ref{splines_manif} introduces the theoretical formulation of the spline interpolation problem on a Riemannian manifold and highlights its connection with kriging. Section~\ref{main_result} presents the main contribution, our finite-element approximation of the spline predictors, while Section~\ref{theory_fe} provides the theoretical foundations leading to this formulation. Section~\ref{sec_anis} describes a method for incorporating local anisotropies. Section~\ref{sec_results} reports numerical results on two analytical test cases, defined on the sphere and the cylinder, as well as on a global pollution study. Finally, Section~\ref{conclu} concludes the paper with a summary and outlines potential directions for future work.

\section{Theory of splines on manifolds}\label{splines_manif}

\subsection{RKHS on a compact manifold}\label{rkhs_sec}

Let $\manif$ be a smooth compact manifold of dimension $d$, equipped with a Riemannian metric $g$, and let $-\Delta$ denote the Laplace-Beltrami operator on $(\manif, g)$. The integral of a function $f$ over $\manif$ is denoted by $\int_\manif f d\mug$, where $d\mug$ is the canonical measure associated to $(\manif, g)$ and $\partial \manif$ denotes its boundary. The spaces $\lm$ and $\hm$ denote, respectively, the set of square-integrable functions and the Sobolev space of order 2. The following assertions are valid for the closed and the Neumann eigenvalue problems. The Spectral Theorem provides that~\cite{Craioveanu2001}:
\begin{itemize}
\item The spectrum of $-\Delta$ is an infinite and countable sequence of real values 
$$0 \leq \lambda_0 \leq \lambda_1 \leq \dots \leq \lambda_k \leq .. $$
in which each eigenvalue appears as many times as its multiplicity. 
\item The eigenvalues $\left(\lambda_k\right)_{k\in \mathbb{N}}$ have finite multiplicity $m_{k}$, and the associated eigenspaces $\left(\mathcal{E}_{\lambda_k}\right)_{k\in \mathbb{N}}$ are orthogonal vector spaces of $L^2(\manif)$.
\item There exists $(\phi_k)_{k\in\mathbb{N}}$ an orthonormal basis of $\lm$ such that $\forall k \in \mathbb{N}, \phi_k \in \mathcal{C}^\infty(\manif)$ is an eigenfunction associated to the eigenvalue $\lambda_k.$ Then, $\forall u \in \lm, u = \sum_{k\in\mathbb{N}} u_k \phi_k$ with $$u_k = \langle \phi_k, u \rangle_{\lm} = \int_{\manif}u \phi_kd\mug.$$
\end{itemize}

As $\manif$ is compact and connected, the eigenvalue $\lambda_0 = 0$ corresponds to the constant functions~\cite{urakawa1993geometry}. Then, we have $m_0 = 1,$ and $\phi_0 = \frac{1}{\int_\manif d\mug}.$ Note that with Dirichlet boundary conditions, the above properties from the Spectral Theorem still hold, but $\lambda_0 > 0$, which largely simplifies the challenges addressed in this study.

Let us now introduce the following subspaces of $\hm,$ 
\begin{equation*}
\raisebox{4pt}{$\Bigg\{$} 
\begin{aligned}
\Vs_0 &= \mathcal{E}_{\lambda_0} ~&&\text{ with associated inner product } \langle u^{(0)}, v^{(0)}\rangle_{0} = u_0v_0 \\
\Vs_1 &= \underset{k\geq 1}{\bigoplus}~ \mathcal{E}_{\lambda_k}~&&\text{ with associated inner product } \langle u^{(1)}, v^{(1)} \rangle_{1} = \sum_{k\geq 1}\lambda_k^2 u_{k}v_k.
\end{aligned}
\end{equation*}
We define the space $\Vs = \Vs_0 \oplus \Vs_1 = \hm$ equipped with the inner product $\langle u, v \rangle = \langle u^{(0)}, v^{(0)} \rangle_{0} + \langle u^{(1)}, v^{(1)} \rangle_{1}$ where $u = u^{(0)}+u^{(1)}$ and $v = v^{(0)}+v^{(1)} \in \Vs$ are the unique decompositions of $u,v \in \Vs$ into their components in $\Vs_0$ and $\Vs_1.$ It can be shown that $\Vs_0, \Vs_1$, and $\Vs$ are RKHS with respective reproducing kernels~\cite{kim2001splines} 
\begin{align}
K_0(\sbf_1, \sbf_2) &= \phi_0(\sbf_1)\phi_0(\sbf_2)\\  K_1(\sbf_1,\sbf_2) &= \sum_{k \geq 1} \frac{1}{\lambda_k^2} \phi_k(\sbf_1)\phi_k(\sbf_2)\label{eq_K1}\\ 
K(\sbf_1,\sbf_2) &= K_0(\sbf_1,\sbf_2)+K_1(\sbf_1,\sbf_2).
\end{align}
These eigenfunctions $\left(\phi_k\right)_{k\in\mathbb{N}}$ play a key role in defining the solution to the spline interpolation problem, detailed in the next section.

\subsection{Spline interpolation on a compact manifold}\label{splines_sec}

Let us consider $n$ observations denoted by $\ybf = \left(y_i\right)_{i=1}^n \in \mathbb{R}^n$,  at points $\Scal = \left(\sbf_i\right)_{i=1}^n$ with $\sbf_i \in \manif, 1\leq i \leq n.$
The spline interpolation problem on $\manif$ is to find $u \in \Vs$ minimizing~\cite{keller2019thin,duchamp2003spline}

\begin{equation}
E(u) \eqdef  \int_{\manif} \lvert\Delta u\rvert^2 d\mug
 = \lVert u \rVert_1 ^2
\end{equation}
subject to the constraint

$$u(\sbf_i) = y_i, 1\leq i \leq n.$$
Here, $\lVert u \rVert_1^2 = \sum_{k\geq 1}\lambda_k^2 \lvert u_k \rvert^2.$ Note that $\lVert \cdot \rVert_1$ is the norm in $\Vs_1$ associated with the inner product $\langle \cdot,\cdot\rangle_1$, but is only a semi-norm in $\Vs$ as $\forall u_0 \in \Vs_0, \lVert u_0 \rVert_1 = 0.$ The solution to this spline interpolation problem is given by Theorem~\ref{theo_wahba}.

\begin{theorem}[Interpolating splines~\cite{wahbabook}]\label{theo_wahba}
If the set of points $\mathcal{S}$ is such that the matrix $\mathbf{K}_1 = \left[K_1(\mathbf{s}_i,\mathbf{s}_j)\right]_{1 \leq i,j \leq n}$ is positive definite, then the solution $u^{\star}$ to the minimization problem is given, for all $\mathbf{s} \in \mathcal{M}$, by
\begin{equation}
u^\star(\sbf) = \azero\phi_0(\sbf) + \bff^\top \kbf_1(\sbf),
\end{equation}
where $ \azero \in \mathbb{R} $, $ \bff = (b_i)_{i=1}^n \in \mathbb{R}^n $, and $\kbf_1(\sbf) = \left(K_1(\sbf_1,\sbf),\dots,K_1(\sbf_n,\sbf)\right)^\top.$
The coefficients $ \azero $ and $ \bff $ satisfy the system
\begin{align}
 \bff^\top\phibfn &= 0,\quad \text{with } \phibfn = \left(\phi_0(\sbf_1),\dots,\phi_0(\sbf_n)\right)^\top\label{eq1} \\
 \azero\phibfn +\Kbf_1\bff &= \ybf,\quad \text{with } \Kbf_1 = \left[K_1(\sbf_i,\sbf_j)\right]_{1\leq i,j\leq n}\label{eq2}
\end{align}
\end{theorem}

The solution $u^\star$ given by the theorem can be made explicit, since it is straightforward to verify that Equations~\eqref{eq1} and \eqref{eq2} are equivalent to:
\begin{align}
\azero &=  \left((\phibfn)^\top\Kbf_1^{-1}\phibfn\right)^{-1}(\phibfn)^\top\Kbf_1^{-1} \ybf\label{eq_a0}\\
\bff &= \Kbf_1^{-1}\left(\ybf-\azero\phibfn\right)\label{eq_b}.
\end{align}
It leads to 
\begin{equation}
u^\star_0(\sbf) = \azero\phi_0(\sbf) + \kbf_1(\sbf)^\top\Kbf_1^{-1}\left(\ybf-\azero\phibfn\right)
\end{equation}
with $\azero$ given in equation~\eqref{eq_a0}. As shown in Appendix \ref{k1_invers}, $\Kbf_1$ is positive definite if and only if $\operatorname{span}\{\wbf_k\}_{k > 0} = \mathbb{R}^n,$ where $\wbf_k = \left(\phi_k(\sbf_1),\dots,\phi_k(\sbf_n)\right)^\top.$ This estimator is known as universal kriging~\cite{armstrong1984problems}, and it also arises as the posterior mean of a Gaussian process regression model~\cite{williams2006gaussian}:

\begin{equation}\label{post_exp}
u^{\star}_0(\sbf) = \mathbb{E}\left(Y_0(\sbf) \mid Y_0(\sbf_1) = \ybf_1,\dots,Y_0(\sbf_n) = \ybf_n\right),
\end{equation}
where $Y_0$ is a Gaussian process of a priori mean $\azero\phi_0$ and covariance kernel $K_1.$ 

\subsection{Smoothing splines on a compact manifold}

The interpolation constraint may become irrelevant when the observed data are noisy. In such cases, the condition can be relaxed, leading to the smoothing splines problem, which seeks to find $u\in V$ minimizing 

\begin{equation}\label{smooth_prob}
\sum_{i=1}^{n} \left(y_i-u(\sbf_i)\right)^2 + \noise^2 \lVert u \rVert_1 ^2.
\end{equation}

\begin{theorem}[Smoothing splines~\cite{wahbabook}]\label{theo_smoothing}

The optimal smoothing splines are defined as

\begin{equation}
u^\star_\noise(\sbf) = \anoise\phi_0(\sbf) + \kbf_1(\sbf)^\top\left(\Kbf_{1}+ \noise^2 \Ibf_{n}\right)^{-1}\left(\ybf-\anoise\phibfn\right),
\end{equation}
with 

\begin{equation}\label{azero_noisy}
\anoise = \left((\phibfn)^\top\left(\Kbf_1 +\noise^2\Ibf_n\right)^{-1}\phibfn\right)^{-1}(\phibfn)^\top\left(\Kbf_1 +\noise^2\Ibf_n\right)^{-1} \ybf.
\end{equation}
\end{theorem}

This solution can again be seen as the posterior mean of a Gaussian process regression model:

\begin{equation}\label{post_exp_2}
u^\star_\noise(\sbf) = \mathbb{E}\left(Y_\noise(\sbf)\mid Y_\noise(\sbf_1)+\noise \eps_1 = \ybf_1,\dots,Y_\noise(\sbf_n)+\noise \eps_n = \ybf_n\right),
\end{equation}
where $Y_\noise$ denotes a Gaussian process of a priori mean $\anoise\phi_0$ and covariance kernel $K_1,$ and $\eps_1,\dots,\eps_n$ are independent standard Gaussian variables. 

Working with a Gaussian process defined on $ \manif $ with covariance kernel $ K_1 $ is then the main objective of this work. The central challenge lies in the formulation of the kernel $ K_1 $, which depends on the spectral decomposition of the Laplace-Beltrami operator $ -\Delta $.  
On the sphere, for example, it is well known that the eigenfunctions of $-\Delta$ are the spherical harmonics~\cite{dai2013spherical}, which enables spline interpolation on this specific manifold, as studied in numerous works~\cite{hitczenko2012some, traas1987smooth,keller2019thin, dunitz2023thin, bonabifard2022semi}. However, for a general manifold, the eigenvalues and eigenfunctions of $ -\Delta $ are not available in closed form. This limitation motivates the present work, which relies on a discrete approximation of the Laplace-Beltrami operator. In the next section, we present the finite-element approximation of the spline predictors. 

\section{Finite element approximation of spline predictors}\label{main_result}

Here, we present a finite element approximation of $u^\star_\noise$, with $\tau = 0$ corresponding to interpolating splines and $\tau > 0$ to smoothing splines, following the approach of~\cite{Lindgren}. This section focuses on the main results, while all theoretical justifications are provided in Section~\ref{theory_fe}.

The manifold $ \manif $ is discretized using $m $ nodes $ \cbf_1, \dots, \cbf_m \in \manif $,  forming a triangulation denoted by $ \mathrm{T} $. A family of compactly supported basis functions $ \left( \psi_j \right)_{j=1}^{m} $ is then defined over $ \manif $. Each function $ \psi_j $ is piecewise linear with respect to the triangulation $ \mathrm{T} $, taking the value $ 1 $ at node $ \cbf_j $ and $ 0 $ at all other nodes, corresponding to the first-order finite element method. Along with these basis functions, we classically have 
\begin{itemize}
\item The $m\times m$ mass matrix $\Mbf$ with $\left[\Mbf\right]_{ij} = \langle \psi_i,\psi_j \rangle_{\lm}$,
\item The $m \times m$ stiffness matrix $\Fbf$ with $\left[\Fbf\right]_{ij} = \langle \nabla \psi_i,\nabla \psi_j \rangle_{\lm}$,
\item The $n\times m$ projection matrix between the observation points and the triangulation nodes, $\Abf_n$, with $\left[\Abf_n\right]_{i,j} = \psi_j(\sbf_i).$ 
\end{itemize}

It can be shown that $\Mbf$ is a symmetric positive definite matrix and $\Fbf$ a symmetric positive semi-definite
matrix~\cite{pereira:tel-02499376}. Let $\sqrt{\Mbf}\in \mathbb{R}^m$ be a matrix such that $\sqrt{\Mbf}\sqrt{\Mbf}^\top = \Mbf$, obtained for instance with the Cholesky decomposition~\cite{benoit1924note}, and now define the matrix $\Sbf$ by

\begin{equation}\label{eq_S}
\Sbf = \left(\sqrt{\Mbf}\right)^{-1}\Fbf\left(\sqrt{\Mbf}\right)^{-\top}.
\end{equation}
It is important to note that the choice of the basis functions $\left(\psi_j\right)_{j=1}^{m}$ leads to sparse matrices $\Mbf$ and $\Fbf$, which facilitates computations when $m$ is large. Here, as is often the case in practice, the matrix $\Mbf$ is replaced by its mass lumping approximation~\cite{quarteroni2016modellistica}, defined as the diagonal matrix with entries $\left[\Mbf\right]_{jj} = \langle \psi_j, 1 \rangle$ for $1 \leq j \leq m$. For simplicity, we denote this mass lumping approximation by $\Mbf$ in the remainder of the text. This choice facilitates straightforward access to the inverse of $\sqrt{\Mbf}$ and ensures that $\Sbf$ is sparse as well.

Finally, for any $\alp > 0,$ we introduce the matrix 

\begin{equation}\label{q_alp}
\tilde{\qbf}_\alp = \sqrt{\Mbf}\Sbf^2\sqrt{\Mbf}^\top + \frac{1}{\alp}\left(\Mbf\phibf\right)\left(\Mbf\phibf\right)^\top,
\end{equation}
where $\phibf = \left(\phi_0(\cbf_j)\right)_{j=1}^{m} = \frac{\mathbf{1}_m}{\lVert \sqrt{\Mbf}^\top \mathbf{1}_m \rVert_2}$ (see Appendix~\ref{appendix_phi0}). Note that the term $\sqrt{\Mbf}\Sbf^2\sqrt{\Mbf}^\top$ is sparse due to the mass-lumping approximation, while the term $\frac{1}{\alpha}\left(\Mbf\phibf\right)\left(\Mbf\phibf\right)^\top$ is a rank-one update involving the known vector $\Mbf\phibf$.  
Using these matrices, Theorem~\ref{theorem_main_result} provides the expression of a finite element approximation of the spline predictors defined in the previous section.

\begin{theorem}[Finite element approximation of spline predictors.]\label{theorem_main_result}
For any $\alpha > 0$, the finite element approximation of the spline predictors is defined, at the triangulation nodes, as the vector
\begin{equation}\label{relation_augment}
\ubf^{\star}_\noise 
= \unoisealp(\ybf) 
+ \Bigg(
   \frac{\phibf - \hh\!\left[\unoisealp\!\left(\Abf_n \phibf\right)\right]}
        {\left(\Mbf\phibf\right)^\top \unoisealp\!\left(\Abf_n \phibf\right)}
   - \phibf 
   + \hh\!\left[\unoisealp\!\left(\Abf_n \phibf\right)\right]
  \Bigg) 
  \left(\Mbf\phibf\right)^\top \unoisealp(\ybf),
\end{equation}
where
$$\unoisealp(\xbf) = \tilde{\qbf}_\alp^{-1} \Abf_n^\top \left( \Abf_n \tilde{\qbf}_\alp^{-1} \Abf_n^\top +\noise^2 \Ibf_n \right)^{-1}\xbf,\quad\quad\xbf\in \mathbb{R}^n$$ and
$\hh\!\left[\ubf\right] 
= \frac{\ubf - \phibf\left(\Mbf\phibf\right)^\top \ubf}{1 - \left(\Mbf\phibf\right)^\top \ubf}$ 
\end{theorem}

In particular, $\ubf^{\star}_\noise$ does not depend on the choice of $\alp > 0.$ However, precautions must be taken in practice when selecting $\alp$ to avoid numerical issues, as discussed in Section~\ref{sec_alp}.
Then, if $\unoisealp(\ybf)$ and $\unoisealp(\Abf_n \phibf)$ are known, the vector $\ubf^\star_\noise$ can be obtained through simple matrix multiplications. Therefore, Algorithms~\ref{algo_ualpzero} and~\ref{algo_ualpnoise} detail the computation of $\unoisealp(\xbf)$ for $\xbf \in \mathbb{R}^n$, with the objective of evaluating it at $\xbf = \ybf$ and $\xbf = \Abf_n \phibf$. Each algorithm corresponds to one of the following scenarios, distinguished for computational convenience:
\begin{enumerate}
\item In the first case, referred to as the \emph{first scenario}, the observation points $ \Scal = \left(\sbf_i\right)_{i=1}^{n} $ are included among the triangulation nodes $ \cbf_1, \dots, \cbf_m $, with $n < m.$ This setting is not particularly restrictive in practice, as the triangulation can typically be constructed to incorporate the observation points. The interpolating splines will be computed only in this case, then working with $\tau = 0.$ Let us denote $I = \{i_1,\dots, i_n\}$ the set of indices such that $s_{k} = c_{i_k}$ for $1\leq k \leq n.$ Then, for $1\leq k \leq n$ and $1\leq j \leq m,$
$$
(\Abf_n)_{k,j} =
\begin{cases}
1 & \text{if } j = i_k, \\
0 & \text{otherwise}.
\end{cases}
$$

Then, for any matrix \(\qbf\) and index sets \(I_1, I_2\), we denote by
\(\left[\qbf\right]_{I_1,I_2}\) the submatrix of \(\qbf\) consisting of the rows indexed by \(I_1\) and the columns indexed by \(I_2\). Similarly, for any vector \(\vbf\), we write \(\big[\vbf\big]_{I_1}\) for the entries of
\(\vbf\) with indices in \(I_1\). Algorithm~\ref{algo_ualpzero} details the computation of $\unoisealp(\xbf)$ for this scenario.

\item In the second case (referred to as the \emph{second scenario}), the observation points $ \Scal $ are not included in the set of triangulation nodes $ \cbf_1, \dots, \cbf_m $. Here, we will compute the smoothing splines ($\noise > 0$), as ensuring the invertibility of the matrices involved and computing the inverses may lead to numerical instability. Algorithm~\ref{algo_ualpnoise} details the computation of $\unoisealp(\xbf)$ for this scenario.
\end{enumerate}

\begin{algorithm}[H]
\caption{Computation of $\uzeralp(\xbf)$ for $\xbf \in \mathbb{R}^n$}
\begin{algorithmic}[1]
\Require $\Mbf, \Fbf, \phibf, I, \alpha, \Abf_n,\xbf$
\State $\Sbf \gets \sqrt{\Mbf}^{-1} \, \Fbf \, \sqrt{\Mbf}^{-\top}$
\State $\qbf \gets \sqrt{\Mbf} \, \Sbf^2 \, \sqrt{\Mbf}^{\top}$
\State Compute the Cholesky decomposition of $\left[\tilde{\qbf}_\alpha\right]_{\bar{I},\bar{I}} = \left[\qbf\right]_{\bar{I},\bar{I}} + \frac{1}{\alpha}\left[\Mbf\phibf\right]_{\bar{I}}\left[\Mbf\phibf\right]_{\bar{I}}^\top$ using sparsity of $\left[\qbf\right]_{\bar{I},\bar{I}}$ and a rank-one update
\State Solve $\left[\tilde{\qbf}_\alpha\right]_{\bar{I},\bar{I}}\ombf = \left[\tilde{\qbf}_\alpha\right]_{\bar{I},I}\xbf$
\State \Return $\begin{bmatrix}
[\uzeralp(\xbf)]_{\bar{I}} \\
[\uzeralp(\xbf)]_{I}\end{bmatrix} = \begin{bmatrix}
 \ombf\\ \xbf
\end{bmatrix}$
\end{algorithmic}\label{algo_ualpzero}
\end{algorithm}

\begin{algorithm}[H]
\caption{$\unoisealp(\xbf)$ for $\xbf \in \mathbb{R}^n$}
\begin{algorithmic}[1]
\Require $\Mbf, \Fbf, \phibf, \alpha, \Abf_n, \xbf, \noise > 0$
\State $\Sbf \gets \sqrt{\Mbf}^{-1} \, \Fbf \, \sqrt{\Mbf}^{-\top}$
\State $\qbf \gets \sqrt{\Mbf} \, \Sbf^2 \, \sqrt{\Mbf}^{\top}$
\State Compute the Cholesky of $\noise^2\tilde{\qbf}_\alp + \Abf_n^\top\Abf_n = \noise^2\qbf + \Abf_n^\top\Abf_n + \frac{\noise^2}{\alp}\left(\Mbf\phibf\right)\left(\Mbf\phibf\right)^\top$ using sparsity of $\noise^2\qbf+ \Abf_n^\top\Abf_n$ and a rank-one update
\State Solve $\left(\noise^2\tilde{\qbf}_\alp + \Abf_n^\top\Abf_n\right)\ombf = \Abf_n^\top\xbf$
\State \Return $\unoisealp(\xbf) = \ombf$
\end{algorithmic}\label{algo_ualpnoise}
\end{algorithm}

All the theoretical justifications for these results are provided in the following section.

\section{Theory of finite element splines on manifolds}\label{theory_fe}

As shown in Equation~\eqref{post_exp_2}, the spline predictor can be interpreted as the posterior mean of the Gaussian process $Y_\noise$, with prior mean $a_\noise\phibf$ and covariance kernel $K_1$. We have $\noise = 0$ for the interpolation problem, and $\noise > 0$ for the smoothing problem. However, since $K_1$ depends on the spectral decomposition of the Laplace-Beltrami operator $-\Delta$, a finite element formulation is proposed here.

First, let us decompose $Y_\noise$ as
\begin{equation}\label{def_y}
Y_\noise = a_\noise\phi_{0} + Z,
\end{equation} with $Z$ a centered Gaussian process (GP) defined on $\manif$ with covariance kernel $K_1$ defined in equation~\eqref{eq_K1}, $\anoise$ defined in equation~\eqref{azero_noisy}, and $\phi_0$ the eigenfunction of $-\Delta$ associated with $\lambda_0=0.$ 

\subsection{Discretization of Z}

The kernel $K_1(\sbf_1,\sbf_2) = \sum_{k \geq 1} \frac{1}{\lambda_k^2} \phi_k(\sbf_1)\phi_k(\sbf_2)$ can be written $K_1(\sbf_1,\sbf_2) = \sum_{k \in \mathbb{N}} f(\lambda_k) \phi_k(\sbf_1)\phi_k(\sbf_2)$ for $(s_1,s_2)\in\manif^2$, with $$f(\lambda) =
\begin{cases} 
0, & \text{if } \lambda = 0 \\
\frac{1}{\lambda^2}, & \text{else.}
\end{cases}
$$
Following the approach
of~\cite{pereira_desassis_allard_2022}, we define the discretized field $Z$ through the expansion 
\begin{equation}\label{expansion}
Z(\sbf) = \sum_{k \in \mathbb{N}}  f^{1/2}(\lambda_k)W_k\phi_k
\end{equation}
with $(W_k)_{k\in\mathbb{N}}$ a sequence of independent standard Gaussian variables and $f^{1/2}$ such that $(f^{1/2})^2 = f.$ Using finite elements, the random fields defined on $\manif$ and specifically the field $Z,$ are then approximated as
$$
Z^{(m)}(\sbf) = \sum_{j=1}^{m} Z_j \psi_j(\sbf) = \Abf(\sbf)^\top \Zbf,
$$
where $ Z_j = Z^{(m)}(\cbf_j) $ for $ 1 \leq j \leq m $, $ \Zbf = \left( Z_1,\dots,Z_m \right)^\top$, and $ \Abf(\sbf) = \left( \psi_1(\sbf),\dots,\psi_m(\sbf) \right)^\top $. 

The objective here is to replace the expansion of $ Z $ given in equation~\eqref{expansion} with an analogous expansion for $Z^{(m)},$ based on the eigenvalues $ \lambda_k^{(m)} $ and eigenvectors $ \phi_k^{(m)} $ of $ -\Delta_m $, the Galerkin approximation of $ -\Delta $ on the triangulation $ \mathrm{T} $. The operator $ -\Delta_m $ is defined explicitly in Appendix~\ref{galerk}. Then we want to define accordingly the weights $\Zbf = \left(Z_1,\dots, Z_m\right)^\top$ such that
\begin{equation}\label{exp_discrete}
Z^{(m)}(\sbf) = \Abf(\sbf)^\top\Zbf = \sum_{k=0}^{m-1} f^{1/2}(\lambda_k^{(m)})W_k\phi_k^{(m)}.
\end{equation} 

As shown in~\cite{lang2023galerkin}, the eigenvalues of $-\Delta_m$ are those of the matrix $\Sbf$ defined in Equation~\eqref{eq_S}. As $\Sbf$ is real, symmetric and positive semi-definite, the matrix is diagonalizable and can be written
\begin{equation}
\Sbf = \Vbf\diag\left(\lambda_0^{(m)},\dots, \lambda_{m-1}^{(m)}\right) \Vbf^\top
\end{equation}
where $\Vbf$ is an orthogonal matrix whose columns are the eigenvectors of $\Sbf$. The work of \cite{lang2023galerkin} also provides the following important result.
\begin{prop}
The weight vector $\Zbf = \left(Z_1,\dots, Z_m\right)^\top$ satisfying Equation~\eqref{exp_discrete} is a centered Gaussian vector with covariance matrix
\begin{equation}\label{eq_sigma}
\sigbf = \left(\sqrt{\Mbf}\right)^{-\top} f(\Sbf) \left(\sqrt{\Mbf}\right)^{-1},
\end{equation}
where $f(\Sbf)$ is the matrix function defined from the eigendecomposition of $\Sbf$ as
$$
f(\Sbf) = \Vbf \diag\left(f(\lambda_0^{(m)}),\dots, f(\lambda_{m-1}^{(m)})\right) \Vbf^\top.
$$
\end{prop}

Then, the associated random field $Z^{(m)}(\sbf) = \Abf(\sbf)^\top\Zbf$ is a centered Gaussian process with covariance kernel 
\begin{equation}\label{eq_k1m}
K_1^{(m)}(\sbf_1, \sbf_2) = \sum_{k=0}^{m-1} f(\lambda_k^{(m)})\phi_k^{(m)}(\sbf_1)\phi_k^{(m)}(\sbf_2) = \Abf(\sbf_1)^{\top}\sigbf \Abf(\sbf_2),~~\sbf_1,\sbf_2 \in \manif.
\end{equation} 

From the covariance kernel of $Z^{(m)}$, the classical kriging formulas can be applied to study the discretization of $Y,$ as presented in the next section.

\subsection{Interpolation formulas}\label{krig_form_section}

We investigate 
\begin{equation}
Y_\noise^{(m)} = \amnoise\phi_0^{(m)} + Z^{(m)}, 
\end{equation}
 the finite element discretization of the target Gaussian process $Y$ introduced in Equation~\eqref{def_y}, with weights 
 
 \begin{equation}\label{eq_weights}
 \Ybf_\noise = \left(Y_\noise^{(m)}(\cbf_j)\right)_{j=1}^{m} = \amnoise\phibf+ \Zbf,
 \end{equation} where
 
\begin{equation}\label{eq_amnoise}
 \amnoise = \left(\phibf^\top \Abf_n^\top\left[\Abf_n \sigbf \Abf_n^\top +\noise \Ibf_n\right]^{-1}\Abf_n\phibf\right)^{-1}\phibf^\top \Abf_n^\top\left[\Abf_n \sigbf \Abf_n^\top+\noise \Ibf_n\right]^{-1} \ybf.
\end{equation}

The expression of $\amnoise$ is based on equation~\eqref{eq_a0}, using the fact that $\left(\phi_0^{(m)}(\sbf_1),\dots, \phi_0^{(m)}(\sbf_n)\right)^\top = \Abf_n\phibf$ and $\left[K_1^{(m)}(\sbf_i,\sbf_j)\right]_{1\leq i,j\leq n} =  \Abf_n\sigbf\Abf_n^\top.$

The objective is to investigate the finite element approximation of both interpolating and smoothing splines, which are defined as posterior expectations in Equations~\ref{post_exp} and \ref{post_exp_2}, respectively. More precisely, the approximated solution at the triangulation nodes are
\begin{equation}
\begin{aligned}
  \ubf^{\star}_\noise &\eqdef \mathbb{E}\left(\Ybf_\noise \mid Y_\noise^{(m)}(\sbf_1)+\noise \eps_1= y_1, \dots, Y_\noise^{(m)}(\sbf_n)+\noise \eps_n = y_n\right) \\
  &\:=  \mathbb{E}\left(\Ybf_\noise \mid \Abf_n\Ybf_\noise + \noise \Ebf = \ybf\right)
 \end{aligned}
 \end{equation}
with $\Ebf = (\eps_1,\dots,\eps_n)^\top \sim \mathcal{N}(0,\Ibf_n),$ and $\noise = 0$ for the interpolating splines.  This leads to Proposition~\ref{theorem_finite_element}.

\begin{prop}\label{theorem_finite_element}
The finite-element approximation of the spline interpolation problem (for $\noise = 0$) and of the smoothing splines problem (for $\noise > 0$) is given by
\begin{equation}\label{krig_formula}
 \ubf^{\star}_\noise 
 = \amnoise\phibf 
 + \sigbf \Abf_n^T 
   \left(\Abf_n \sigbf (\Abf_n)^T + \noise^2 \Ibf_n\right)^{-1}
   \bigl(\ybf - \amnoise \Abf_n \phibf\bigr).
\end{equation}
\end{prop}

The inversibility of $\Abf_n\sigbf (\Abf_n)^T$ and $\Abf_n\sigbf (\Abf_n)^T + \noise^2\Ibf_n$ are studied for the first and the second scenario in Appendix~\ref{inverse_mat_krig}. Nevertheless, the kriging formula of equation~\eqref{krig_formula} is not straightforward to exploit in practice. This issue will be addressed in the following section.

\subsection{Intrinsic GMRF}

The main challenge here arises from the fact that the size $m$ of the triangulation must be very large to ensure accurate approximations. In such cases, computing the eigendecomposition of the matrix $\Sbf \in \mathbb{R}^{m \times m}$ becomes impractical, rendering both $f(\Sbf)$ and the covariance matrix $\sigbf$ intractable. To address this issue, a common approach is to work with the precision matrix $\qbf$, that is the inverse of $\sigbf$. In particular, when the function $f$ is the inverse of a polynomial $P$ that remains positive on $\mathbb{R}_+$, the precision matrix admits a convenient expression: $\sigbf^{-1} = \left(\sqrt{\Mbf}\right) P(\Sbf) \left(\sqrt{\Mbf}\right)^{\top},$  which is sparse when using mass lumping. Using this formulation, the approximation $\Ybf_\noise$ defines a GMRF~\cite{rue2005gaussian}, which enables computational simplifications.

However, in our situation, as explained in Appendix~\ref{first_scenario_appendix}, $\sigbf$ has rank $m-1$ and is not invertible, rendering $\Ybf_\noise$ an intrinsic GMRF~\cite{simpson2006sampling}. The idea is to introduce a modified GMRF that has an invertible covariance matrix, inspired by the Bayesian kriging framework~\cite{diggle2002bayesian}, in which the intercept term $\amzero$ is modeled as a random variable $\Amzero$ with a Gaussian prior distribution. More precisely, we denote
\begin{equation}
\tilde{\Ybf}_\alp = \Amzero(\alp)\phibf + \Zbf
\end{equation}
where $\Amzero(\alp)\sim \mathcal{N}(0, \alp)$ independent from $\Zbf.$

\begin{prop}\label{prop_y}$\tilde{\Ybf}_\alp = \Amzero(\alp)\phibf + \Zbf $ has 
\begin{itemize}
\item Non-singular covariance matrix $\tilde{\sigbf}_\alp = \sigbf + \alp \phibf \phibf ^\top$
\item Precision matrix $\tilde{\qbf}_\alp = \sqrt{\Mbf}\Sbf^2\sqrt{\Mbf}^\top + \frac{1}{\alp}\left(\Mbf\phibf\right)\left(\Mbf\phibf\right)^\top$
\end{itemize}
\end{prop}

The proof is provided in Appendix~\ref{appendix_precision}. As previously highlighted, the matrix $ \tilde{\qbf}_\alpha $ is particularly useful in this context, as it consists of two components that facilitate  efficient Cholesky decompositions (see Sections~\ref{first_scenar}~and~\ref{second_scenar}):
\begin{itemize}
  \item The term $ \sqrt{\Mbf}\Sbf^2\sqrt{\Mbf}^\top $ is sparse thanks to the mass-lumping approximation.
  \item The term $ \frac{1}{\alpha}\left(\Mbf\phibf\right)\left(\Mbf\phibf\right)^\top $ is a rank-one update involving the known vector $ \Mbf\phibf $.
\end{itemize}

Then, as stated in Theorem~\ref{theorem_main_result}, using the posterior expectation 
\begin{equation}\label{augment_krig}
\unoisealp(\xbf) \eqdef \mathbb{E}\left(\tilde{\Ybf}_\alp \mid \Abf_n\tilde{\Ybf}_\alp + \noise \mathbf{E} = \xbf\right)=\tilde{\qbf}_\alp^{-1} \Abf_n^\top \left( \Abf_n \tilde{\qbf}_\alp^{-1} \Abf_n^\top +\noise^2 \Ibf_n \right)^{-1}\xbf,
\end{equation}
for $\xbf \in \mathbb{R}^{n}$, $\noise \geq 0$ and $\alp > 0,$ provides the tractable expression of Equation~\eqref{relation_augment} for our finite element approximation $\ubf^{\star}_\noise: $ 
$$\ubf^{\star}_\noise 
= \unoisealp(\ybf) 
+ \Bigg(
   \frac{\phibf - \hh\!\left[\unoisealp\!\left(\Abf_n \phibf\right)\right]}
        {\left(\Mbf\phibf\right)^\top \unoisealp\!\left(\Abf_n \phibf\right)}
   - \phibf 
   + \hh\!\left[\unoisealp\!\left(\Abf_n \phibf\right)\right]
  \Bigg) 
  \left(\Mbf\phibf\right)^\top \unoisealp(\ybf).$$ 
  
The detailed proof is provided in Appendix~\ref{appendix_augmented}. The posterior expectation $\unoisealp(\xbf)$ is tractable, as explained in Section~\ref{first_scenar} for the first scenario ($\noise = 0$) and in Section~\ref{second_scenar} for the second scenario ($\noise > 0$). It can then be evaluated for $\xbf = \ybf$ and $\xbf = \Abf_n \phibf$, after which only simple matrix multiplications remain.

\subsection{First scenario: computation of $\uzeralp(\xbf)$ for $\xbf \in \mathbb{R}^n$ }\label{first_scenar}

Here we detail the computation of $\uzeralp(\xbf)$, for the \emph{first scenario} where the observation points are included among the triangulation nodes. The procedure is detailed in Algorithm~\ref{algo_ualpzero}. This computation is particularly relevant when $\xbf = \ybf$ or $\xbf = \Abf_n \phibf$ in equation~\eqref{relation_augment}. As introduced in Section~\ref{main_result}, let $I = \{i_1, \dots, i_n\}$
denote the set of indices such that \(s_k = c_{i_k}\) for \(1 \leq k \leq n\).
We define its complement as $\bar{I} = \{1, \dots, m\} \setminus I.$ Here, $$\uzeralp(\xbf) = \mathbb{E}\left(\tilde{\Ybf}_\alp \mid \big[\tilde{\Ybf}_\alp\big]_{I} = \xbf\right),$$ and the vector $\begin{bmatrix}
\big[\tilde{\Ybf}_\alpha\big]_{\bar{I}} \\
\big[\tilde{\Ybf}_\alpha\big]_{I}\end{bmatrix}$
has precision matrix 
$
\begin{bmatrix}
\big[\tilde{\qbf}_\alpha\big]_{\bar{I},\bar{I}} & \big[\tilde{\qbf}_\alpha\big]_{\bar{I},I} \\
\big[\tilde{\qbf}_\alpha\big]_{I,\bar{I}} & \big[\tilde{\qbf}_\alpha\big]_{I,I}
\end{bmatrix}.
$

Then, using the classical kriging formula involving the precision matrix, we have 
$$
\begin{cases}
\big[\uzeralp(\xbf)\big]_{\bar{I}} &= \big[\tilde{\qbf}_\alpha\big]_{\bar{I},\bar{I}}^{-1}\big[\tilde{\qbf}_\alpha\big]_{\bar{I},I}\xbf \\
\big[\uzeralp(\xbf)\big]_{I} &= \xbf.
\end{cases}
$$

The vector $\left[\tilde{\qbf}_\alpha\right]_{\bar{I},\bar{I}}^{-1}\left[\tilde{\qbf}_\alpha\right]_{\bar{I},I}\xbf$ can be obtained via the Cholesky decomposition of $\left[\tilde{\qbf}_\alpha\right]_{\bar{I},\bar{I}}$, followed by solving a sparse linear system. More precisely, the Cholesky factorization of $\left[\tilde{\qbf}_\alpha\right]_{\bar{I},\bar{I}} = \left[\sqrt{\Mbf}\Sbf^2\sqrt{\Mbf}^\top\right]_{\bar{I},\bar{I}} + \frac{1}{\alpha}\left[\Mbf\phibf\right]_{\bar{I}}\left[\Mbf\phibf\right]_{\bar{I}}^\top$ exploits the sparsity of $\left[\sqrt{\Mbf}\Sbf^2\sqrt{\Mbf}^\top\right]_{\bar{I},\bar{I}}$ and a rank-one update. Additional details about the sparse linear systems are provided in Appendix~\ref{cholesky_appendix}.

\subsection{Second scenario : computation of $\unoisealp(\xbf)$ for $\xbf \in \mathbb{R}^n$}\label{second_scenar}

This section focuses on the computation of $\unoisealp(\xbf) = \mathbb{E}\left(\tilde{\Ybf}_\alp \mid \Abf_n\tilde{\Ybf}_\alp + \noise \mathbf{E} = \xbf\right)$ with $\noise >0.$
Again, this is required when $\xbf = \ybf$ or $\xbf = \Abf_n \phibf$ in equation~\eqref{relation_augment}. Algorithm~\ref{algo_ualpnoise} summarizes the corresponding steps. \\~\

We have $\unoisealp(\xbf) = \tilde{\qbf}_\alp^{-1} \Abf_n^\top \left( \Abf_n \tilde{\qbf}_\alp^{-1} \Abf_n^\top +\noise^2 \Ibf_n \right)^{-1}\xbf,$ with $\Abf_n \tilde{\qbf}_\alp^{-1} \Abf_n^\top +\noise^2 \Ibf_n $ invertible as $\Abf_n \tilde{\qbf}_\alp^{-1} \Abf_n^\top$ is positive semidefinite:
$\forall \xbf \in \mathbb{R}^n, \xbf^\top \Abf_n \tilde{\qbf}_\alp^{-1} \Abf_n^\top\xbf = \left(\Abf_n^\top\xbf\right)^\top \tilde{\qbf}_\alp^{-1} \Abf_n^\top\xbf \geq 0.$

From Woodbury formula and as detailed in Appendix~\ref{appendix_woodbury}, it can be shown that

\begin{equation}\label{eq_noise_augment}
\tilde{\qbf}_\alp^{-1} \Abf_n^\top\left( \Abf_n \tilde{\qbf}_\alp^{-1} \Abf_n^\top + \noise^2\Ibf_n\right)^{-1}\xbf = \left(\noise^2\tilde{\qbf}_\alp+\Abf_n^\top\Abf_n\right)^{-1}\Abf_n^\top\xbf.
\end{equation}

Since $\left[\Abf_n^\top\Abf_n\right]_{ij} = 0$ whenever $\psi_i(\sbf_k)\psi_j(\sbf_k) = 0$ for all $1 \leq k \leq n$, the matrix $\Abf_n^\top\Abf_n$ is sparse. Then, the Cholesky decomposition of $$\noise^2\tilde{\qbf}_\alp+\Abf_n^\top\Abf_n = \noise^2\sqrt{\Mbf}\Sbf^2\sqrt{\Mbf}^\top+\Abf_n^\top\Abf_n + \frac{\noise^2}{\alp}\left(\Mbf\phibf\right)\left(\Mbf\phibf\right)^\top$$ can be computed by first exploiting the sparsity of $\noise^2\sqrt{\Mbf}\Sbf^2\sqrt{\Mbf}^\top + \Abf_n^\top\Abf_n$, and then applying a rank-one update. Once the Cholesky factor is available, solving sparse linear systems yields 
$$
\left(\noise^2\tilde{\qbf}_\alpha + \Abf_n^\top\Abf_n\right)^{-1} \Abf_n^\top \xbf.
$$

\subsection{Discussion on $\alp$}\label{sec_alp}

Theoretically, all the presented results hold for any $\alpha > 0.$ 
In practice, however, the choice of $\alpha$ is crucial to avoid numerical instabilities. 
If $\alpha$ is chosen too close to $0$, computing the inverse of $\boldsymbol{\Sigma}$ may become numerically unstable or infeasible. On the other hand, as $\alpha \to \infty$, we obtain
$\tilde{\mathbf{Q}}_{\alpha} \approx \sqrt{\mathbf{M}}\,\mathbf{S}^2\,\sqrt{\mathbf{M}}^{\top},$
which is not invertible.

More rigorous arguments can be developed to formalize these intuitions and show that a suitable choice of $\alpha$ satisfies
\[
\frac{1}{\sqrt{\alpha}} \in \big[\lambda_{1}^{(m)},\, \lambda_{m-1}^{(m)}\big],
\]
where $\lambda_{0}^{(m)} \le \lambda_{1}^{(m)} \le \cdots \le \lambda_{m-1}^{(m)}$ denote the eigenvalues of $\Sbf$. Details on selecting such a value of $\alpha$ are provided in Appendix~\ref{appendix_alpha}, where the power iteration method~\cite{mises1929praktische} is employed.

\paragraph*{First Scenario.} Under the mass lumping approximation, the Cholesky decomposition of
\[
\left[\tilde{\qbf}_\alpha\right]_{\bar{I},\bar{I}} 
= \left[\sqrt{\Mbf}\Big(\Sbf^2 + \tfrac{1}{\alpha}(\sqrt{\Mbf}^\top\phibf)(\sqrt{\Mbf}^\top\phibf)^\top\Big)\sqrt{\Mbf}^\top\right]_{\bar{I},\bar{I}}
\]
can be obtained with the Cholesky decomposition of
\[
\left[\Sbf^2 + \tfrac{1}{\alpha}(\sqrt{\Mbf}^\top\phibf)(\sqrt{\Mbf}^\top\phibf)^\top\right]_{\bar{I},\bar{I}}.
\]

Since $\sqrt{\Mbf}^\top\phibf \in \ker(\Sbf)$ (see Appendix~\ref{inverse_mat_krig}), the eigenvalues of $\Sbf^2$ and
\[
\tilde{\Sbf}^2_\alpha = \Sbf^2 + \tfrac{1}{\alpha}(\sqrt{\Mbf}^\top\phibf)(\sqrt{\Mbf}^\top\phibf)^\top
\]
are identical, except that the zero eigenvalue $\left(\lambda_0^{(m)}\right)^2 = 0$ is replaced by $\tfrac{1}{\alpha}.$

To mitigate numerical issues, it is essential to keep the condition number of $\tilde{\Sbf}^2_\alpha$ as small as possible. 
This condition number is defined as
\[
\kappa(\tilde{\Sbf}_\alpha^2) = \frac{\lambda_{\max}(\tilde{\Sbf}_\alpha^2)}{\lambda_{\min}(\tilde{\Sbf}_\alpha^2)}.
\]
where $\lambda_{\max}(\tilde{\Sbf}_\alpha^2)$ and $\lambda_{\min}(\tilde{\Sbf}_\alpha^2)$ 
denote the maximum and minimum eigenvalues of $\tilde{\Sbf}_\alpha^2$, respectively. Hence, the desirable choice is to enforce
\[
\frac{1}{\sqrt{\alpha}} \in \big[\lambda_1^{(m)}, \, \lambda_{m-1}^{(m)}\big].
\]

The largest eigenvalue of a matrix can, for instance, be computed using the power iteration method~\cite{mises1929praktische}, 
which only requires performing matrix--vector products with this matrix. Since $\lambda_1^{(m)}$ is the largest eigenvalue of $\left(\Sbf+\Ibf_{m}\right)^{-1} - \bigl(\sqrt{\Mbf}^\top\phibf\bigr)\bigl(\sqrt{\Mbf}^\top\phibf\bigr)^\top,$
both $\lambda_1^{(m)}$ and $\lambda_{m-1}^{(m)}$ can be computed using the power-iteration method.

\paragraph*{Second scenario.} Analogously, in the second scenario, the Cholesky decomposition of
\(\noise^2 \tilde{\qbf}_\alpha + \Abf_n^\top \Abf_n\)
can be derived from the Cholesky decomposition of
\(\noise^2 \tilde{\Sbf}_\alpha^2 + \sqrt{\Mbf}^{-1} \Abf_n^\top \Abf_n \sqrt{\Mbf}^{-\top}\). 
We also have the following bounds on the eigenvalues~\cite{weyl1912asymptotische}:

\[
\begin{cases}
\lambda_{\max}\!\Big(\noise^2\tilde{\Sbf}_\alpha^2 + \sqrt{\Mbf}^{-1} \Abf_n^\top \Abf_n \sqrt{\Mbf}^{-\top}\Big) 
\le
\lambda_{\max}\!\Big(\sqrt{\Mbf}^{-1} \Abf_n^\top \Abf_n \sqrt{\Mbf}^{-\top}\Big) + \lambda_{\max}(\noise^2\tilde{\Sbf}_\alpha^2), \\[1em]
\lambda_{\min}\!\Big(\noise^2\tilde{\Sbf}_\alpha^2 + \sqrt{\Mbf}^{-1} \Abf_n^\top \Abf_n \sqrt{\Mbf}^{-\top}\Big) 
\ge 
\lambda_{\min}\!\Big(\sqrt{\Mbf}^{-1} \Abf_n^\top \Abf_n \sqrt{\Mbf}^{-\top}\Big) + \lambda_{\min}(\noise^2\tilde{\Sbf}_\alpha^2).
\end{cases}
\]

As a consequence, the condition number satisfies

\[
\kappa\Big(\tilde{\Sbf}_\alpha^2 + \sqrt{\Mbf}^{-1} \Abf_n^\top \Abf_n \sqrt{\Mbf}^{-\top}\Big)
\le 
\frac{\lambda_{\max}\!\Big(\sqrt{\Mbf}^{-1} \Abf_n^\top \Abf_n \sqrt{\Mbf}^{-\top}\Big) + \noise^2\lambda_{\max}(\tilde{\Sbf}_\alpha^2)}
{\lambda_{\min}\!\Big(\sqrt{\Mbf}^{-1} \Abf_n^\top \Abf_n \sqrt{\Mbf}^{-\top}\Big) + \noise^2\lambda_{\min}(\tilde{\Sbf}_\alpha^2)}.
\]

Choosing \(\frac{1}{\sqrt{\alpha}} \in [\lambda_1^{(m)}, \, \lambda_{m-1}^{(m)}]\) is thus again a relevant choice in this context. \\~\

Introducing $\tilde{\Ybf}_\alpha$ with a carefully chosen $\alpha$ provides a tractable formulation of the finite-element solution. Nevertheless, this solution is isotropic, as evidenced by the covariance kernel $K_1^{(m)}$ (see equation~\eqref{eq_k1m}). This limitation may lead to unsatisfactory results when the underlying phenomenon exhibits preferred directions, a situation addressed in Section~\ref{sec_anis}.

\section{Anisotropic splines}\label{sec_anis}

\subsection{Local anisotropy through space deformation}

The idea behind constructing solutions with local anisotropies is to introduce a deformation of the manifold $\manif$ embedded in $\mathbb{R}^d$, through its Riemannian metric $g$, as presented in~\cite{pereira:tel-02499376}. More precisely, for any $\sbf \in \manif$, we consider a coordinate chart $(U, x)$ containing $\sbf$, where $U$ is an open subset of $\manif$ and $x : U \to \mathbb{R}^d$ is a homeomorphism mapping $U$ to an open subset of $\mathbb{R}^d$. The metric is then defined as
\begin{equation}
g_\sbf(\ubf_\sbf, \vbf_\sbf) = \left(\ubf_\sbf^x\right)^\top \Gbf^x(\sbf) \, \vbf_\sbf^x,
\end{equation}
with $\ubf_\sbf^x$ and $\vbf_\sbf^x$ the representative vectors of $\ubf_\sbf$ and $\vbf_\sbf$ of the tangent space $T_\sbf \manif$ at $\sbf$, with respect to the chart $(U,x).$ The matrix $\Gbf^x(\sbf)$ is the representative of the Riemannian metric at $\sbf$ in these coordinates and is symmetric positive definite. It can be interpreted as a deformation matrix, and admits the following diagonalization:
$$
\Gbf^x(\sbf) = \Rbf^x(\sbf) \, \Dbf^x(\sbf)^2 \, \left(\Rbf^x(\sbf)\right)^\top,
$$
where $\Rbf^x(\sbf)$ is an orthogonal matrix and $\Dbf^x(\sbf)$ is diagonal with positive entries. With this Riemannian metric, $Y^{(m)}_\noise$ becomes locally isotropic after applying the linear change of variables $\hbf \mapsto \Dbf^x(\sbf) \left(\Rbf^x(\sbf)\right)^\top \hbf$, where $\hbf$ denotes an infinitesimal displacement vector around $\sbf$.

For $d \in \{2,3\}$, the matrix $\Gbf^x(\sbf)$ admits a natural geometric interpretation. The diagonal entries of $\Dbf^x(\sbf)$ can be interpreted as local scaling factors $\rho_1^x(\sbf), \dots, \rho_d^x(\sbf)$ along the principal directions defined by the chart $(U, x)$, while $\Rbf^x(\sbf)$ represents a rotation matrix parameterized by angles $\ang_j^x(\sbf)$, denoted by

$$\Rbf_{\ang^x(\sbf)} = 
\begin{pmatrix}
\cos\ang^x(\sbf) & -\sin\ang^x(\sbf) \\
\sin\ang^x(\sbf) & \cos\ang^x(\sbf)
\end{pmatrix} $$
when $d=2,$ and 
{\setlength{\arraycolsep}{1pt} 
\begin{align*}
\Rbf_{\ang_1^x(\sbf),\,\ang_2^x(\sbf),\,\ang_3^x(\sbf)} &= \\
\begin{pmatrix}
\cos\ang_3^x(\sbf) & -\sin\ang_3^x(\sbf) & 0 \\
\sin\ang_3^x(\sbf) & \cos\ang_3^x(\sbf) & 0 \\
0 & 0 & 1
\end{pmatrix}
&\begin{pmatrix}
\cos\ang_2^x(\sbf) & 0 & -\sin\ang_2^x(\sbf) \\
0 & 1 & 0 \\
\sin\ang_2^x(\sbf) & 0 & \cos\ang_2^x(\sbf)
\end{pmatrix}
\begin{pmatrix}
1 & 0 & 0 \\
0 & \cos\ang_1^x(\sbf) & -\sin\ang_1^x(\sbf) \\
0 & \sin\ang_1^x(\sbf) & \cos\ang_1^x(\sbf)
\end{pmatrix}
\end{align*}
}
when $d=3.$

This spatial deformation, represented through the Riemannian metric, affects the computation of the integral over $\manif$, and consequently influences the mass and stiffness matrices, $\Mbf$ and $\Fbf$ and finally the Bayesian precision matrix $\tilde{\qbf}_\alp$ (see Proposition~\ref{prop_y}). Further details are provided in the Supplementary Material~\cite{charliesire_2025} for the specific cases of the cylindrical and spherical surfaces, which are 2-dimensional Riemannian manifolds embedded in $\mathbb{R}^3$. These cases allow for the introduction of rotations and scaling within the local charts defined by the 2D cylindrical coordinates $(\theta, z)$ or spherical coordinates $(\theta, \varphi)$, to obtain interpretable deformations. To estimate the optimal hyperparameters for the matrix $\Gbf^x(\sbf)$, the idea here is to use maximum likelihood estimation.

\subsection{Maximum likelihood estimation}\label{mle}

In theory, local anisotropies can be introduced by considering fully defined anisotropy fields on the manifold. In practice, however, a manageable number of parameters $\param$ must be specified to define these fields. The issue of incorporating local anisotropies is discussed further in Section~\ref{real_world}. In simpler settings, however, the hyperparameters of $\Gbf^x(\sbf)$, namely the rotation angles and scaling factors, are assumed to be constant across the manifold $\manif$. In that case, all hyperparameters to be estimated can be collected into the vectors

\begin{itemize}
\item $\param = \left(\ang, \rho_1, \rho_2\right)$ if $d=2$
\item $\param = \left(\ang_1, \ang_2, \ang_3, \rho_1, \rho_2,\rho_3\right)$ if $d=3$.
\end{itemize}

These hyperparameters influence $\sigbf$, $\tilde{\qbf}_\alp$, $\phibf$, and $\amnoise$ in the following, but this dependency is not explicitly indicated for simplicity. 

The objective is to optimize the likelihood of the observations based on our statistical model. 
Recall that our finite-element solution is the conditional expectation of the field 
$Y_\noise^{(m)} = \Abf(s)^\top \Ybf_\noise$, with $\Ybf_\noise$ defined in 
equation~\eqref{eq_weights}. 
Then, the Gaussian field at the observation points 
$\bigl[Y_\noise^{(m)}(\sbf_1), \dots, Y_\noise^{(m)}(\sbf_n)\bigr]^\top 
= \Abf_n \Ybf_\noise$ 
is Gaussian with mean $\amnoise \Abf_n \phibf$ and covariance matrix 
$\Kbf_\noise = \Abf_n \sigbf \Abf_n^\top + \noise^2 \Ibf_n$, 
where $\noise = 0$ in the \emph{first scenario} and $\noise > 0$ in the \emph{second scenario}.

The log-likelihood is then 
\begin{equation}
\llik_\noise(\param) =  -\frac{1}{2} \left(n\log(2\pi)+\log \lvert \Kbf_\noise \rvert + \left(\ybf - \amnoise \Abf_n \phibf\right)^\top \Kbf_\noise^{-1}\left(\ybf - \amnoise \Abf_n \phibf\right) \right),
\end{equation}
The computation of the likelihood is split into two terms, the quadratic form $\ell^{(1)}_\noise(\param)$ and the log-determinant $\ell^{(2)}_\noise$:
\begin{itemize}
\item $\ell^{(1)}_\noise(\param) = \left(\ybf - \amnoise \Abf_n \phibf\right)^\top \left(\Abf_n \sigbf \Abf_n^\top+\noise^2\Ibf_n\right)^{-1}\left(\ybf - \amnoise\Abf_n \phibf\right)$
\item $\ell^{(2)}_\noise=\log \lvert \Abf_n \sigbf \Abf_n^\top+ \noise^2\Ibf_n\rvert,$
\end{itemize}

with details provided in Appendix~\ref{likeli}. Note that the modified Gaussian vector $\tilde{\Ybf}_\alp$ does not appear in our statistical model. 
It was introduced to enable the computation of the conditional expectation of $Y_\noise^{(m)}$, 
and will again be useful here for the computation of the likelihood. 

An optimization algorithm is then required to maximize the objective function $\param \mapsto \llik_\noise(\param)$. In this work, we use the CMA-ES algorithm~\cite{hansen2006cma} to estimate the optimal hyperparameters, a robust gradient-free optimization algorithm designed to efficiently explore the search space and escape local minima. In the second scenario, the observation noise can either be fixed a priori or estimated by optimizing $(\noise, \param) \mapsto \llik_\noise(\param).$

\section{Application}\label{sec_results}

The method is evaluated on two different compact manifolds. It is worth noting that the geometry of the manifold poses no inherent limitation, provided that a triangulation is available. The first test case is the sphere, which is particularly relevant since it allows comparison with classical approaches based on spherical harmonics. The second test case is the surface of a cylinder, which is useful in many contexts, such as studies involving tubes~\cite{aiba1981heat,macher2013heated}. For both the sphere and the cylinder, we investigate analytical problems. For the sphere, we also consider real-world data through the analysis of carbon dioxide fluxes. All test cases considered here exhibit smooth variations, as spline predictors are specifically designed for phenomena where such smoothness is assumed a priori. To illustrate the advantages of splines in this setting, a comparison with a Mat\'ern random field is provided in Appendix~\ref{matern} for the pollution dataset. However, the main focus of this article is the construction and implementation of spline-based predictions on manifolds rather than an exhaustive benchmarking study, and it should be noted that spline predictors are not well suited to rough fields, for which other covariance kernels may be more appropriate.

\subsection{Splines on the sphere}

As previously mentioned, on the sphere, a basis $\left(\phi_k\right)_{k\in\mathbb{N}}$ of eigenfunctions of $-\Delta$ can be identified with the spherical harmonics~\cite{dai2013spherical}. Following~\cite{keller2019thin}, working with splines on the sphere is equivalent to considering a Gaussian field with covariance kernel  
\begin{equation}
C(\sbf_1, \sbf_2) = \sum_{k=0}^{+\infty} \frac{2k+1}{k^2(k+1)^2} \, P_k\!\left(\cos\big(\xi(\sbf_1,\sbf_2)\big)\right),
\end{equation}
where $P_k$ is the Legendre polynomial of degree $k$, and $\xi(\sbf_1,\sbf_2)$ denotes the spherical angle between $\sbf_1$ and $\sbf_2$.  

To obtain a more tractable expression for this kernel, one may consider the truncated expansion  
\begin{equation}
C_K(\sbf_1, \sbf_2) = \sum_{k=0}^{K} \frac{2k+1}{k^2(k+1)^2} \, P_k\!\left(\cos\big(\xi(\sbf_1,\sbf_2)\big)\right).
\end{equation}
The truncated kernel $C_K$ can then be directly incorporated into the classical kriging formula. In this work, we set $K = 40$, following the recommendations of~\cite{keller2019thin}.

\subsubsection{Analytical function}\label{analytical_sphere}

The analytical function $ f_\text{sphere} $ investigated here is defined as
$$
\begin{array}{ccccc}
f_\text{sphere} & : & [0, \pi] \times [0, 2\pi] & \to & \mathbb{R} \\
  &   & (\theta, \phi)             & \mapsto & \cos\left(2\theta + \phi + \frac{\pi}{4}\right) \sin^2(\theta)
\end{array}
$$
where $\theta$ and $\phi$ are the spherical coordinates.
\begin{figure}
\centering
\includegraphics[width=0.7\textwidth]{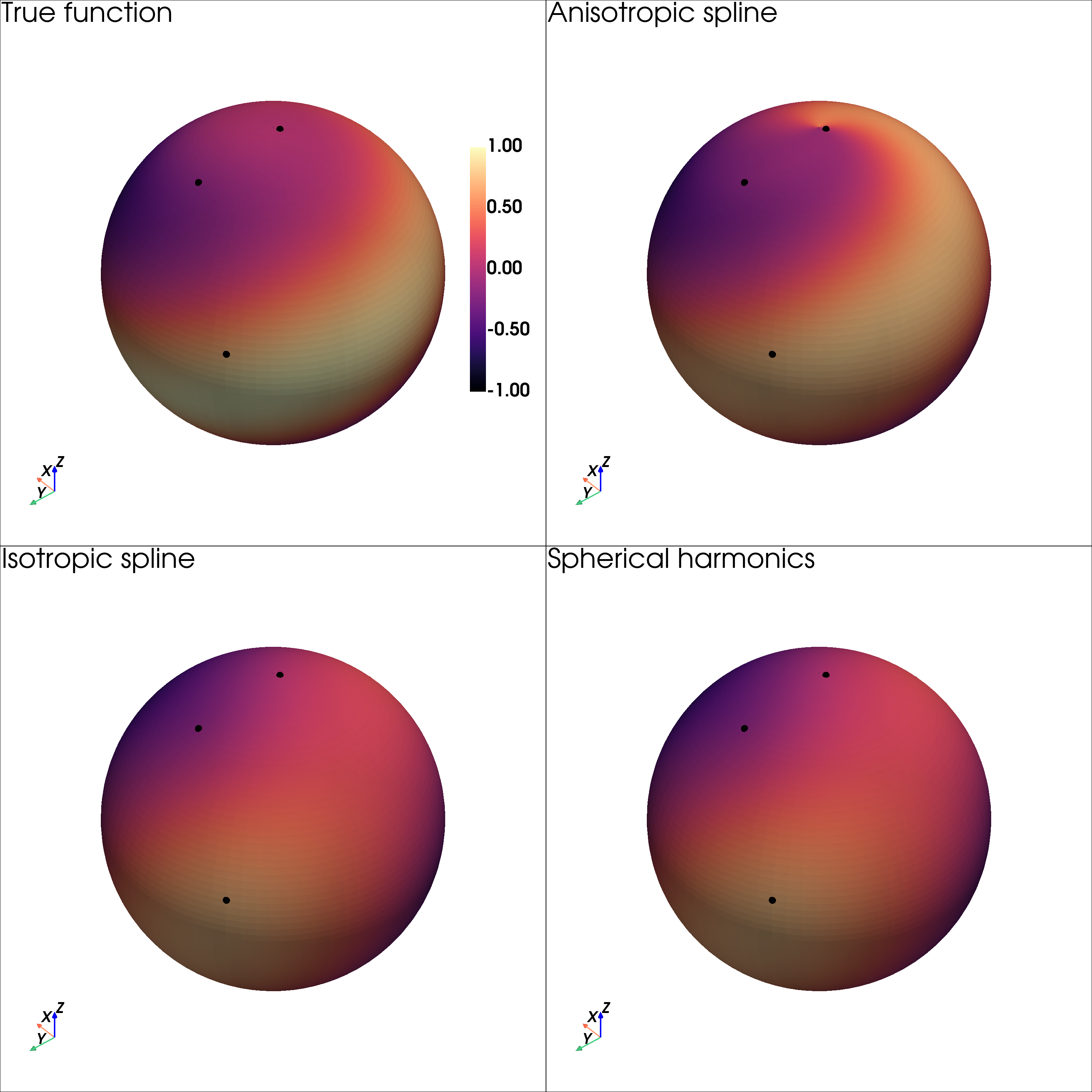}
\caption{Results for the analytical function on the sphere in the first scenario. $n=10$ observation points are shown as black dots. 
Top left: true function. 
Top right: prediction with anisotropies induced in the local charts defined by spherical coordinates $(\theta, \phi)$. 
Bottom left: isotropic splines (no space deformation). 
Bottom right: splines using a kernel based on spherical harmonics.}
\label{results_sphere}
\end{figure}

A triangulation with $ m = 6400 $ nodes is constructed on the unit sphere. Then, $ n = 10 $ observation points are selected as follows. First, a Maximin Latin Hypercube Sampling design~\cite{kenny2000algorithmic}, denoted $\mathbb{X}$, is generated in the spherical coordinate space $(\theta,\phi)$. In the first scenario, the observation points are constrained to lie on the triangulation mesh, so the final design is obtained by selecting the mesh nodes nearest to the points in $\mathbb{X}$. In the second scenario, the final design is exactly $\mathbb{X}$ itself, and the observation noise $\noise >0$ is considered. Here, only the results for the first scenario (i.e., the interpolation problem) are presented, while the results for the second scenario are very similar and are provided in the Supplementary Material~\cite{charliesire_2025}. We emphasize that the problem of finding the optimal design of observations on the sphere is not addressed here, as our focus lies on the prediction method. Regarding the anisotropies, they are optimised using $\param = \left(\ang, \rho_1, \rho_2\right),$ representing a constant rotation and two constant scaling factors in the local charts of spherical coordinates $(\theta, \phi).$ 

\begin{figure}[h]
\centering
\includegraphics[width=0.7\textwidth]{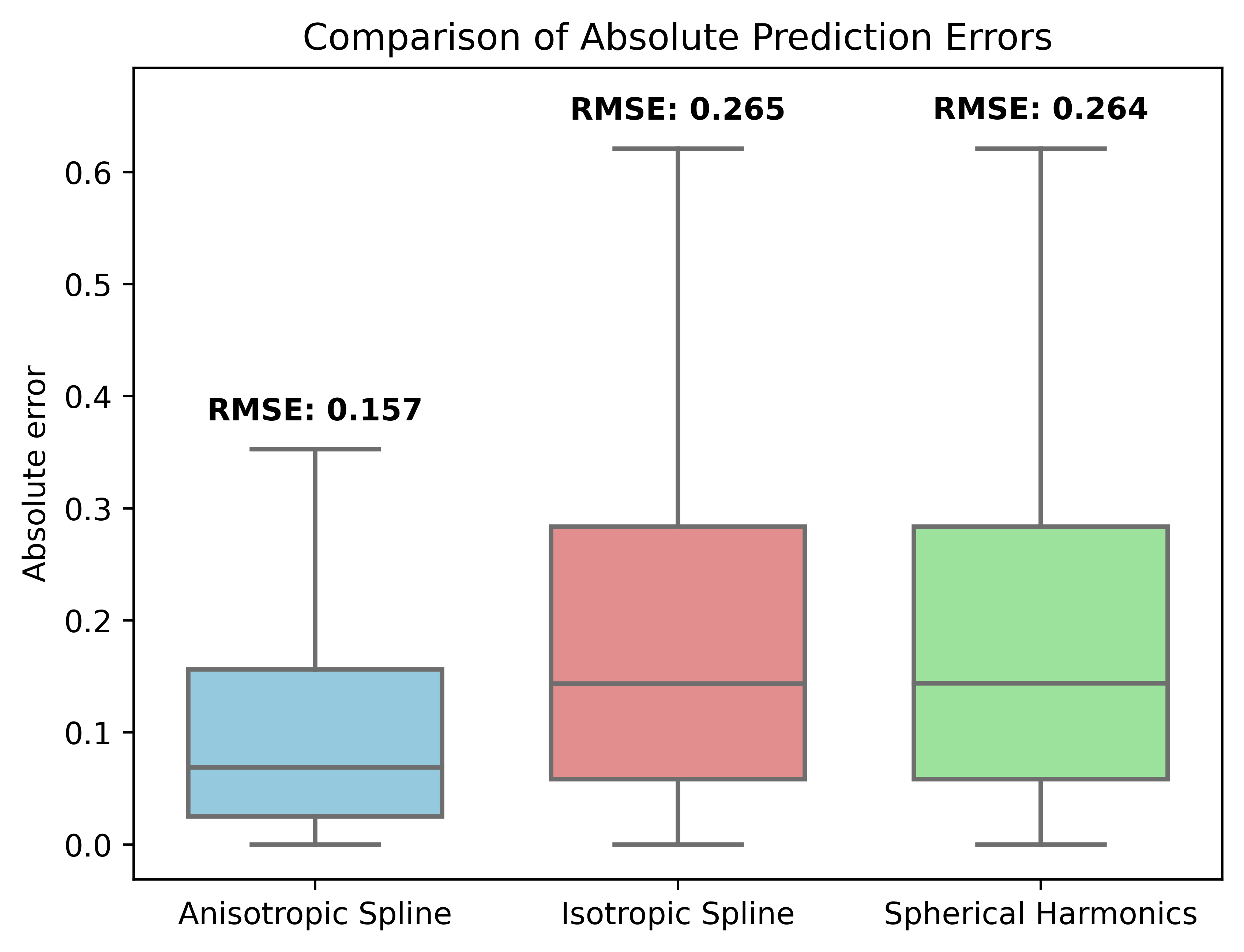}
\caption{Boxplots of absolute prediction errors at the triangulation nodes for the analytical function on the sphere. 
Each boxplot corresponds to a different prediction method: isotropic splines, anisotropic splines, and spherical harmonics.}
\label{boxplots_sphere}
\end{figure}

The results are shown in Figure~\ref{results_sphere}. The isotropic splines produced by our method yield results very similar to those obtained with spherical harmonics, which is promising for the effectiveness of the approach. 
Moreover, introducing anisotropies proves particularly relevant here, enabling a better capture of the function's behavior. 
Two-dimensional representations in spherical coordinates are provided in Appendix~\ref{appendix_2d} to facilitate the visualization of predictions across the entire domain. A comparison of the distribution of the corresponding prediction errors, further summarized by the root mean square error (RMSE), is provided in Figure~\ref{boxplots_sphere}.

\begin{figure}
\centering
\includegraphics[width=0.7\textwidth]{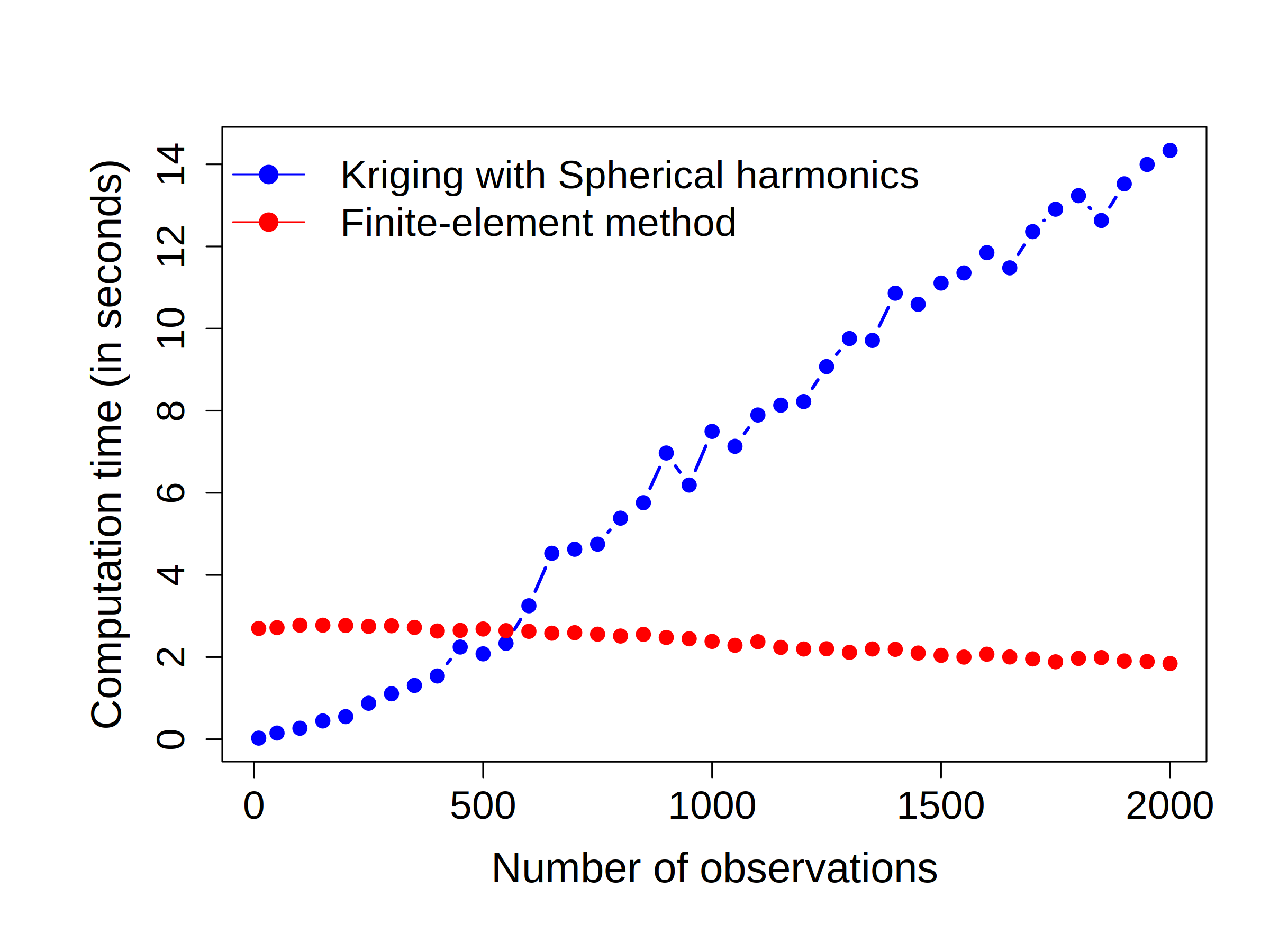}
\caption{Computation time of predictions at the triangulation points as the number of observation points increases from $10$ to $2000$. 
Red: our finite-element-based method. 
Blue: classical kriging with spherical harmonics.}
\label{time_duration}
\end{figure}

Note that, as in the other application test cases, the number of observation points $n$ considered here is small, to highlight the benefit of accounting for anisotropy which is naturally less pronounced when $n$ is very large, except in the presence of highly localized anisotropies. Nevertheless, one of the main advantages of our method, based on the finite element approximation, is precisely that it can accommodate a large number of observations. As emphasized in Algorithm~\ref{algo_ualpzero}, the influence of $n$ on the computational cost is negligible, since the main cost lies in the Cholesky decomposition of $\left[\tilde{\qbf}_\alpha\right]_{\bar{I},\bar{I}}$. This point is illustrated in Figure~\ref{time_duration}, which reports the computation time of predictions at the triangulation points as the number of observation points increases, and compares the performance of the classical approach based on spherical harmonics with that of our method. The figure clearly shows that the classical strategy becomes increasingly expensive as $n$ grows. Moreover, in classical kriging, an additional observation noise must be introduced once $n$ reaches several hundreds in order to compute the inverse of the covariance matrix, an issue that does not arise with our method.

\subsubsection{Real-world data}\label{real_world}

To work with real-world data, we used the greenhouse gas reanalysis dataset 
provided by the Copernicus Atmosphere Monitoring Service (CAMS)~\cite{cams_egg4_2021}. 
Further details about this dataset can be found in~\cite{agusti2023cams}. 
In this study, we focus on carbon dioxide at the 50~hPa pressure level 
on December 16, 2020. A grid in spherical coordinates $(\theta,\phi)$ with a step size of $1.5^\circ$ is first constructed, and a triangulation is built from these nodes. Then, a set of $n=50$ observation points is selected following the procedure described earlier. To account for local variations in anisotropy across the sphere, we model the anisotropies as a continuous field rather than a global constant. This field is parameterized by a restricted set of variables to ensure that the maximum log-likelihood estimation remains computationally tractable. Specifically, we focus on the spatial evolution of the rotation angles $\alpha^{x}(\mathbf{s})$ within the local spherical coordinate chart $(\theta, \phi)$.

To maintain methodological consistency with our proposed spline framework on $\mathcal{M}$, and assuming that rotation angles vary smoothly over the manifold, we define the field $\left(\ang^{x}(\mathbf{s})\right)_{\mathbf{s} \in \mathcal{M}}$ using spline interpolation. This approach allows us to reconstruct the full field from a sparse set of control points:

\begin{itemize}
    \item Pilot points: We introduce a reduced grid of $r$ pilot points $\{\tilde{\mathbf{s}}_i\}_{i=1}^{r}$ on the manifold $\mathcal{M}$.
    \item Optimization variables: The rotation parameters at these specific locations, 
    $\param = \bigl(\ang^{x}(\tilde{\mathbf{s}}_{i})\bigr)_{i=1}^{r}$, 
    are treated as the variables to be inferred through log-likelihood optimization.
    \item Field reconstruction: Leveraging our contribution to spline interpolation, the local rotation $\ang^{s}(\mathbf{s})$ is extended to any $\mathbf{s} \in \mathcal{M}$ by interpolating the values defined at the pilot points.
\end{itemize}

In the present work, we take $r=25$, while the scaling factors are kept constant, with $\rho_{1}=2$ and $\rho_{2}=1/2$. This choice emphasizes the direction given by $\ang^{x}(\sbf)$ while limiting the number of  parameters in the log-likelihood optimization. This construction of local anisotropies is proposed as an original approach to an open and challenging problem, and  provide a structured way to handle spatial non-stationarity while keeping the parameter space manageable. However, this approach entails certain trade-offs and limitations, which are discussed in detail in Section~\ref{conclu}, along with potential alternatives.

\begin{figure}[h]
\centering
\includegraphics[width=0.7\textwidth]{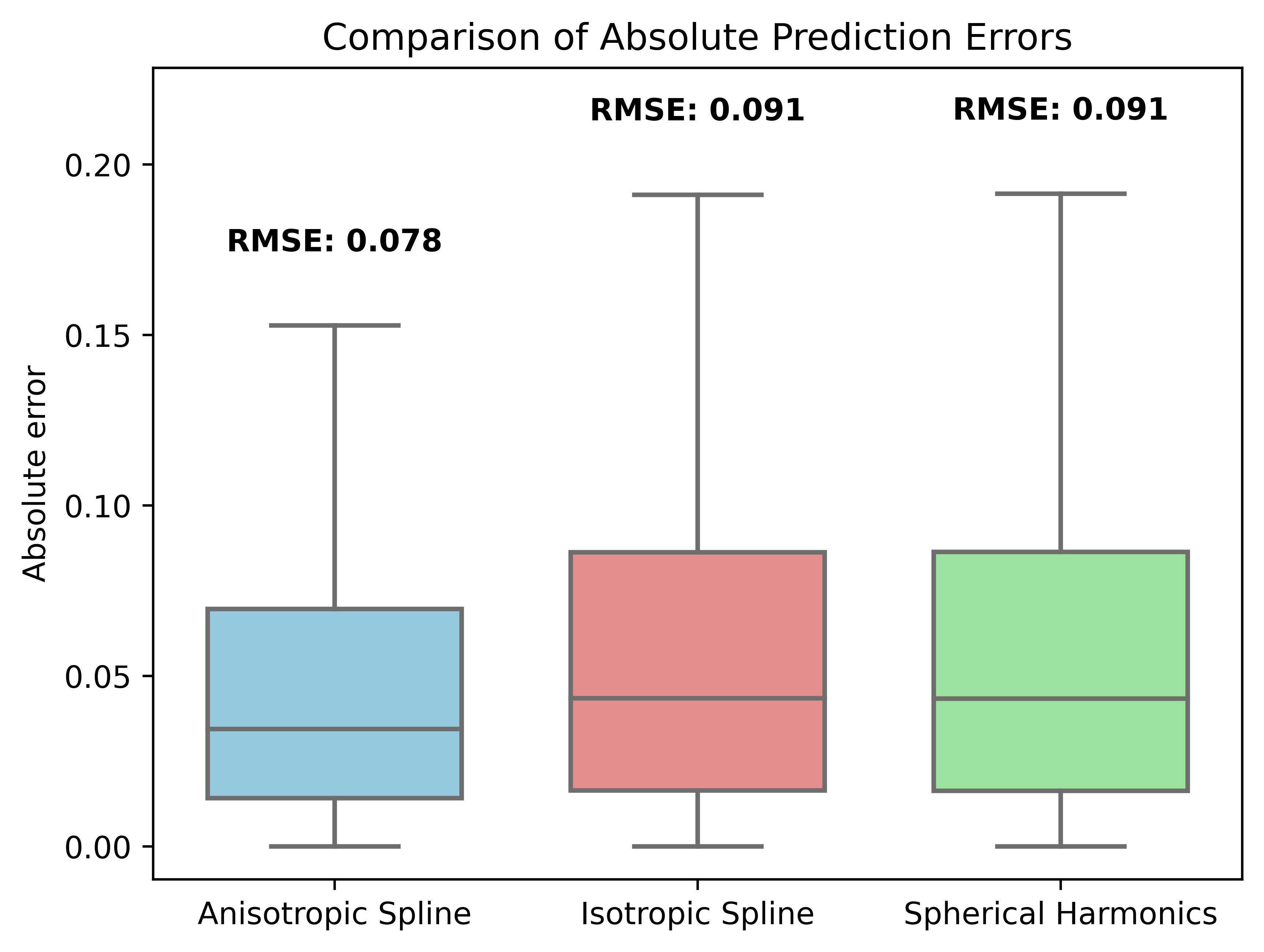}
\caption{Boxplots of absolute prediction errors at the triangulation nodes for the real-world data on the sphere. 
Each boxplot corresponds to a different prediction method: isotropic splines, anisotropic splines, and spherical harmonics.}
\label{boxplots_earth}
\end{figure}

The results are shown in Figure~\ref{results_earth} in 3D for the first scenario, and 2D results are provided in Appendix~\ref{appendix_2d}. The carbon dioxide concentration is expressed here as a mass fraction (kg of CO$_2$ per kg of moist air) and subsequently normalized to the interval $[0,1]$. Once again, the isotropic splines produced by our method yield results that are very close to those obtained with spherical harmonics, thereby confirming the effectiveness of the approach. Furthermore, the introduction of anisotropies proves particularly valuable, as it enables a more accurate representation of the phenomenon's behavior, as supported by the RMSE reported in Figure~\ref{boxplots_earth}.

\begin{figure}
\centering
\includegraphics[width=0.7\textwidth]{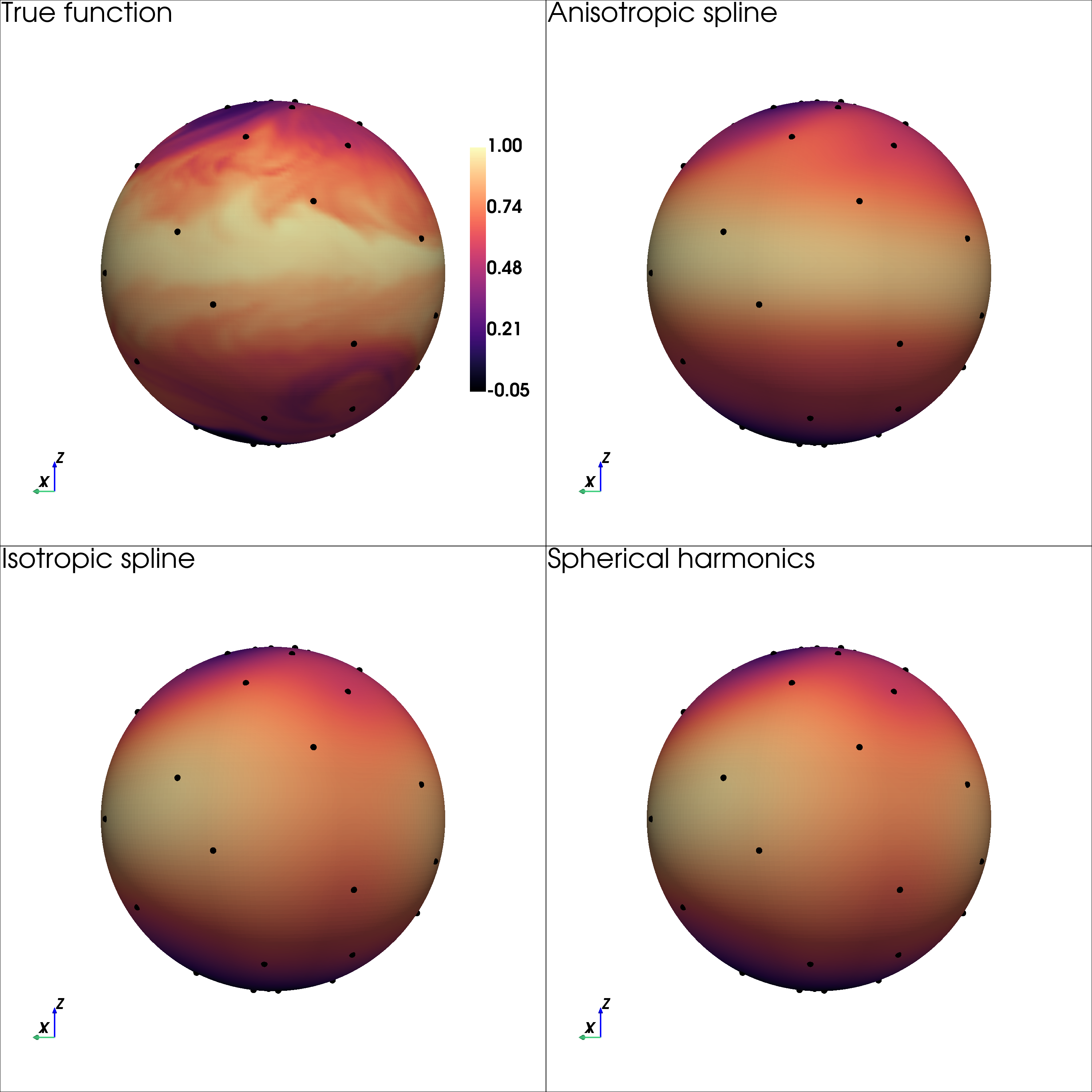}
\caption{Results for the normalized concentration of carbon dioxide on the sphere in the first scenario. $n = 50$ observation points are shown as black dots. 
Top left: true function. 
Top right: prediction with anisotropies induced in the local charts defined by spherical coordinates $(\theta, \phi)$. 
Bottom left: isotropic splines (no space deformation). 
Bottom right: splines using a kernel based on spherical harmonics.}
\label{results_earth}
\end{figure}

\subsection{Splines on the cylinder surface}

The analytical function $ f_\text{cyl} $ investigated here is defined as
$$
\begin{array}{ccccc}
f_{\mathrm{cyl}} & : & [0, 2\pi] \times [z_{\min}, z_{\max}] & \longrightarrow & \mathbb{R} \\
& & (\theta, z) & \longmapsto & \text{exp}\left[-\frac{3}{4}\lVert p\left(\theta, z\right) \rVert ^2\right]
\end{array}
$$

with $p\left(\theta, z\right) = \Pbf
\begin{bmatrix}
\cos\theta \\[0.5em]
z/z_{\max}
\end{bmatrix}$,  $\Pbf
= \begin{bmatrix}
0.5 & 0 \\
0 & 1.5
\end{bmatrix}
\begin{bmatrix}
\cos\frac{\pi}{5} & -\sin\frac{\pi}{5} \\
\sin\frac{\pi}{5} &  \cos\frac{\pi}{5}
\end{bmatrix},$ and $\theta$ and $z$ the cylindrical coordinates.

We consider a cylinder of radius \(1\), with \(z_{\min} = 0\) and \(z_{\max} = 10\). A triangulation is constructed from a grid of points in the \((\theta, z)\) space, chosen so that the resulting triangles in 3D space are isosceles, ensuring a regular mesh. This results in \(m = 70 \times 115 = 8050\) nodes in total. The design of observations is built similarly to the procedure described in Section~\ref{analytical_sphere}. The anistropies are defined by a vector $\param = \left(\ang, \rho_1, \rho_2\right),$ representing a constant rotation and two constant scaling factors in the local charts of cylindrical coordinates $(\theta, z).$ 
\begin{figure}
\centering\includegraphics[width=1\textwidth]{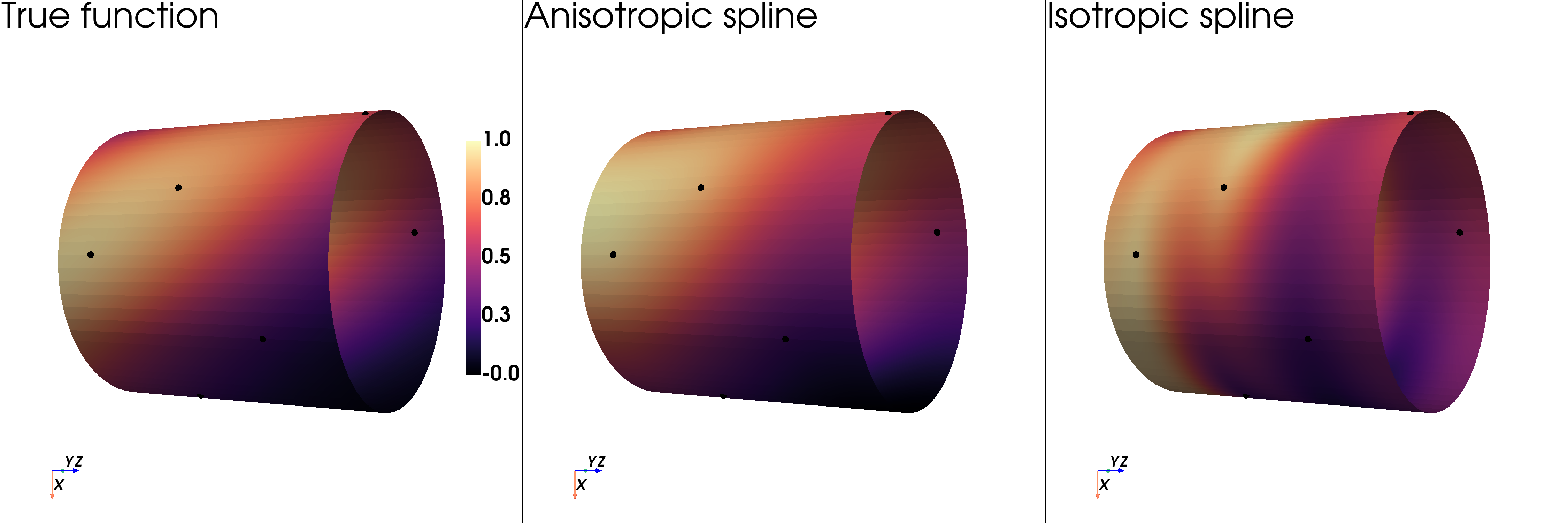}
\caption{Results for the analytical function on the cylinder in the first scenario. $n=10$ observation points are shown as black dots. 
Left: true function. 
Middle: prediction with anisotropies induced in the local charts defined by cylindrical coordinates $(\theta, z)$. 
Right: isotropic splines (no space deformation).}
\label{results_cylinder}
\end{figure}

\begin{figure}[h]
\centering
\includegraphics[width=0.7\textwidth]{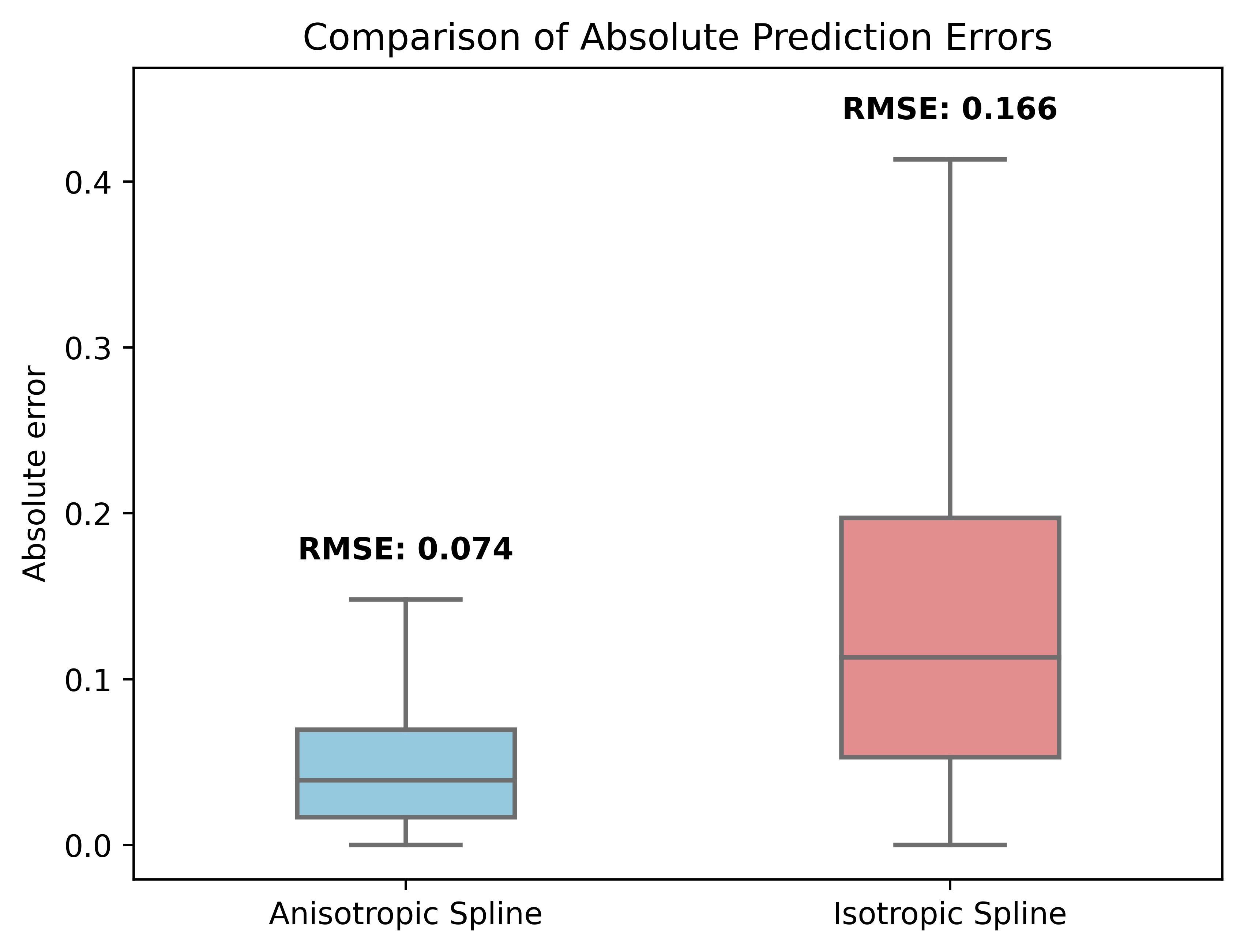}
\caption{Boxplots of absolute prediction errors at the triangulation nodes for the analytical function on the cylinder. 
Each boxplot corresponds to a different prediction method: isotropic and anisotropic splines.}
\label{boxplots_cylinder}
\end{figure}

The results for the first scenario are shown in Figure~\ref{results_cylinder}.  
As before, the results for the second scenario are presented in the Supplementary Material~\cite{charliesire_2025} and are very similar.  
For visualization purposes, the radius has been increased to better illustrate the behavior of the functions. Two-dimensional plots in the space of the cylindrical coordinates are provided in Appendix~\ref{appendix_2d} to facilitate interpretation of the predictions over the entire domain. Again, the anisotropies introduced here appear particularly effective at capturing the behavior of the true function, yielding very promising results, as highlighted by the lower prediction errors of Figure~\ref{boxplots_cylinder}.

\section{Summary and perspectives}\label{conclu}

This article introduces a novel approach for constructing interpolating or smoothing splines on a general compact, connected and orientable manifold \(\manif\). The proposed prediction method yields smooth estimates within the domain, since it arises from a minimization problem (under constraints in the case of interpolating splines) involving an energy functional closely related to the Laplace--Beltrami operator \(-\Delta\) on \(\manif\). Although the analytical solution has been well established since the pioneering work of Wahba~\cite{wahbabook}, it is not computationally tractable in practice, as it requires access to the spectrum of \(-\Delta\). In this work, we exploit the equivalence between this solution and the Gaussian process regression problem, which can be addressed using finite-element approximations on a triangulated mesh of \(\manif\), as thoroughly investigated in~\cite{pereira_desassis_allard_2022}. An important advantage of this approach is its ability to construct solutions with local anisotropies by deforming the spatial domain \(\manif\), thereby producing non-stationary solution. Nevertheless, classical developments encounter difficulties due to the emergence of a singular covariance matrix in our context. To overcome this issue, we propose an augmented GMRF with uncertainty on the trend. This formulation yields a sparse precision matrix that is easy to compute, facilitating calculations, particularly via sparse linear systems. A relation between the kriging formulas based on the augmented GMRF and the initial spline-based solution is then established. 

The method is evaluated on analytical toy problems defined on the sphere and the cylinder, as well as on real-world data on Earth. The case of the sphere is particularly interesting, since the spectrum of the Laplace--Beltrami operator is known, with the spherical harmonics serving as eigenvectors. In this setting, interpolating or smoothing splines can be computed without relying on a finite-element approximation. The results show identical solutions between this approach and our proposed method, which strongly supports the validity of the results. Moreover, the introduction of anisotropies enabled by the finite-element approximation proves highly effective in accurately capturing the behavior of the functions, as highlighted by the prediction errors displayed in Figures~\ref{boxplots_sphere}, \ref{boxplots_earth} and \ref{boxplots_cylinder}. \\~\

Although the study presented here is comprehensive, several avenues for further development remain. First, this work aims to provide a formal basis for spline interpolation on manifolds, offering a framework and extended in future studies. Our current focus has been on the interpolation problem itself, which is a deterministic approach providing a single predicted value. However, incorporating prediction uncertainty is essential in many scientific applications. In this framework, such uncertainty arises naturally through the equivalence with universal kriging, but exploiting the universal kriging variance is a non-trivial task here, but it can be estimated by generating conditional simulations. A specific challenge arises as the a priori GMRF $\Ybf_{\noise}$ does not have an invertible covariance matrix. This issue of simulating intrinsic GMRFs can be addressed by simulating $\tilde{\mathbf{Y}}_{\alpha}$ using its invertible precision matrix, followed by a correction step to obtain the target simulations~\cite{bolin2021efficient}. Another promising alternative would be to use Takahashi recursions~\cite{takahashi1973formation, zammit2018sparse} to compute posterior variances, though the singularity of the covariance matrix would again require careful treatment.

Another possible line of study concerns the solution of sparse linear systems. 
In our approach, these systems are solved using Cholesky decompositions, 
as described in Appendix~\ref{cholesky_appendix}. 
This choice is particularly relevant since the computation of matrix determinants, 
required for the log-likelihood estimation (see Appendix~\ref{likeli}), 
becomes straightforward once the Cholesky factorization is available. 
Nevertheless, alternative techniques could also be considered for solving the linear systems. For example, the conjugate gradient method~\cite{hestenes1952methods} 
may be more appropriate when the size~$m$ of the triangulation is too large 
to allow for Cholesky decompositions.
 
Finally, a significant advantage of the proposed method lies in its capacity to handle anisotropic splines. While the results are particularly promising for phenomena exhibiting constant anisotropy, we have also introduced a strategy to account for local anisotropies. The primary challenge in this context is defining a complete anisotropy field while maintaining a parsimonious parameterization. To address this, we define a field of rotation angles via spline interpolation based on $r$ values at specific pilot points, which are subsequently tuned through log-likelihood optimization. Although this strategy provides a practical framework for incorporating local anisotropies, its scope remains focused on the broader objective of establishing spline operations on manifolds. A comprehensive investigation of local anisotropy deserves dedicated future research. For instance, while the pilot points in this study were arranged on a grid, their number and placement could be further refined using physical constraints or active learning procedures, which could allow for placing these points sequentially. Furthermore, physical data, such as wind vectors in environmental modeling, could directly inform the rotation angles. Future work could also compare our approach with alternative strategies, such as defining the anisotropy field as a linear combination of basis functions~\cite{ingebrigtsen2014spatial} or using vector fields derived from the gradients of real-valued sinusoidal functions~\cite{fuglstad2015exploring}.

\section*{Declarations}

\begin{itemize}
\item Funding: This work was fully supported by the Chaire Geolearning funded by Andra, BNP-Paribas, CCR and SCOR Foundation.
\item Competing interests: The authors have no competing interests to declare that are relevant to the content of this article.
\item Data availability: The data and code used to generate the results presented in this article are provided in the Supplementary Material~\cite{charliesire_2025}.
\end{itemize}

\bibliography{References}

\begin{appendices}

\section{Positive definitiveness of $\Kbf_1.$}\label{k1_invers}

For $k \in \mathbb{N}^\star,$ let us denote 
$\wbf_k = \left(\phi_k(\sbf_1),\dots,\phi_k(\sbf_n)\right)^\top.$

\begin{align*}
\forall \xbf \in \mathbb{R}^n,~~\xbf^\top \Kbf_1 \xbf &= \sum_{i=1}^n \sum_{j=1}^n  x_i x_j \sum_{k>0} \frac{1}{\lambda_k^2} \phi_k(\sbf_i) \phi_k(\sbf_j) \\
&= \sum_{k>0}  \frac{1}{\lambda_k^2} \sum_{i=1}^n \sum_{j=1}^n  x_i x_j \phi_k(\sbf_i) \phi_k(\sbf_j) \\
&= \sum_{k>0}  \frac{1}{\lambda_k^2} \left\lVert \sum_{i=1}^n  x_i \phi_k(\sbf_i) \right \rVert ^2
\end{align*} 
Then, \begin{align*}
\xbf^\top \Kbf_1 \xbf = 0 &\Leftrightarrow  \forall k >0, \sum_{i=1}^n  x_i \phi_k(\sbf_i)=0 \\
&\Leftrightarrow \forall k >0, \langle \xbf, \wbf_k\rangle_{\mathbb{R}^n} = 0
\end{align*}
Finally, it comes

$$\Kbf_1\text{ is positive definite} \Leftrightarrow \operatorname{span}\{\wbf_k\}_{k > 0} = \mathbb{R}^n.$$
\section{Galerkin approximation}\label{galerk}

This section defines the Galerkin approximation $-\Delta_m$ of the Laplace-Beltrami opperation $-\Delta.$ Let us denote $V_m$ the linear span of $\psi_1,\dots,\psi_m$ the piecewise linear functions associated with the triangulation $\tri.$ $-\Delta_m$ is the endomorphism mapping any $f \in V_m$ to $-\Delta_m f,$ with 

$$\forall u \in V_m, \langle -\Delta_m f, u \rangle_{\lm} = \langle - \Delta f, u \rangle_{\lm}.$$

\section{Computation of $\phibf$}\label{appendix_phi0}

By definition, the function $\phi_0^{(m)}$ has unit norm in $\lm$, that is,
\begin{equation*}
\left\langle \sum_{k=1}^{m} [\phibf]_k \psi_k,\ \sum_{k=1}^{m} [\phibf]_k \psi_k \right\rangle_{\lm} = 1,
\end{equation*}
which implies that 
\begin{equation*}
\phibf^\top \Mbf \phibf = \left( \sqrt{\Mbf}^\top \phibf \right)^\top \sqrt{\Mbf}^\top \phibf = 1.
\end{equation*}

Since $\phibf$ is constant, it follows that
\begin{equation}
\phibf = \frac{\mathbf{1}_m}{\lVert \sqrt{\Mbf}^\top \mathbf{1}_m \rVert_2}. 
\end{equation}

\section{Inversibility of the covariance matrices}\label{inverse_mat_krig}

This section investigates the inversibility of $\Abf_n\sigbf (\Abf_n)^T$ and $\Abf_n\sigbf (\Abf_n)^T + \noise^2\Ibf_n$ for the \emph{first scenario} and the \emph{second scenario}, respectively.

\subsection{First scenario}\label{first_scenario_appendix}

We know that $ f(\Sbf) $ is positive semidefinite of rank $ m - 1 $, since for all $ 1 \leq i \leq m - 1 $, we have $ f(\lambda_i^{(m)}) > 0 $, and $ f(\lambda_0^{(m)}) = 0 $. Therefore, the matrix
$
\sigbf = \left(\sqrt{\Mbf}\right)^{-\top} f(\Sbf) \left(\sqrt{\Mbf}\right)^{-1}
$
is also positive semidefinite of rank $ m - 1 $. 

Now let us identify a vector in $\text{Ker}(\sigbf).$ By definition of the basis function $\left(\psi_j\right)_{j=1}^{m},$ we have $\forall \sbf \in \manif, \sum_{j=1}^{m} \psi_j(\sbf) = 1,$ and then 
$$\forall \sbf \in \manif, \sum_{j=1}^{m} \nabla\psi_j(\sbf) = 0.$$

As $\phibf$ is constant, $\Sbf \left(\sqrt{\Mbf}\right)^{\top}\phibf = \left(\sqrt{\Mbf}\right)^{-1}\Fbf\left(\sqrt{\Mbf}\right)^{-\top}\left(\sqrt{\Mbf}\right)^{\top}\phibf = \left(\sqrt{\Mbf}\right)^{-1}\Fbf\phibf = 0.$

Then, $\left(\sqrt{\Mbf}\right)^{\top}\phibf \in \text{Ker}(\Sbf)\subset \text{Ker}(f(\Sbf)).$

Finally, $$\sigbf\Mbf\phibf = \left(\sqrt{\Mbf}\right)^{-\top} f(\Sbf) \left(\sqrt{\Mbf}\right)^{-1}\sqrt{\Mbf}\left(\sqrt{\Mbf}\right)^\top\phibf = 0.$$

This yields a vector in $ \ker(\sigbf) $, and since $ \dim(\ker(\sigbf)) = 1 $, it follows that
$$
\text{for any } \xbf \in \mathbb{R}^m, \quad \xbf \in \ker(\sigbf) \;\Leftrightarrow\; \xbf \propto \Mbf\phibf = \frac{\Mbf \mathbf{1}_m}{\left\lVert \left(\sqrt{\Mbf}\right)^\top \mathbf{1}_m \right\rVert_2}.
$$

This implies that every component of $ \xbf \neq 0$ is non-zero:  
$$
\xbf \in \ker(\sigbf)\setminus \{0\} \;\Rightarrow\; \forall 1 \leq i \leq m,\ \xbf_i \neq 0.
$$

As a reminder, $ I = \{i_1, \dots, i_n\} $ denotes the set of indices such that $ s_k = c_{i_k} $ for $ 1 \leq k \leq n $.

Let $ \ybf \in \mathbb{R}^n \setminus \{0\} $. We define a vector $ \tilde{\ybf} \in \mathbb{R}^m $ by
$$
\tilde{\ybf}_i =
\begin{cases}
\ybf_k & \text{if } i = i_k \text{ for some } 1 \leq k \leq n, \\
0 & \text{otherwise}.
\end{cases}
$$

Then,
$$
\ybf^\top \Abf_n \sigbf \Abf_n^\top \ybf = \tilde{\ybf}^\top \sigbf \tilde{\ybf}.
$$
Since $\ybf \neq 0$, there exists at least one index $i_k$ such that $\tilde{\ybf}_{i_k} \neq 0$, hence $ \tilde{\ybf} \notin \ker(\sigbf) $, and therefore $ \tilde{\ybf}^\top \sigbf \tilde{\ybf} > 0 $. This shows that the matrix $ \Abf_n \sigbf \Abf_n^\top $ is positive definite.

\subsection{Second scenario}

For the \emph{second scenario}, the covariance matrix $ \Abf_n \sigbf \Abf_n^\top + \noise^2 \Ibf_n $ is invertible if and only if $ -\noise^2 $ is not an eigenvalue of $ \Abf_n \sigbf \Abf_n^\top $.

As $\sigbf$ is positive semidefinite, $$\forall \xbf \in \mathbb{R}^{m}\setminus \{0\}, \xbf^\top\Abf_n \sigbf \Abf_n^\top\xbf =  \left(\Abf_n^\top\xbf\right) \sigbf \Abf_n^\top\xbf \geq 0.$$

Then, $\Abf_n \sigbf \Abf_n^\top$ is positive semidefinite, and $-\noise^2$ is not an eigenvalue of $ \Abf_n \sigbf \Abf_n^\top .$

\section{Precision matrix of $\tilde{\Ybf}_\alp$}\label{appendix_precision}

$\tilde{\Ybf}_\alp = \Amzero(\alp)\phibf + \Zbf $ with 
\begin{itemize}
\item $\Zbf$ of covariance matrix $\sigbf = \left(\sqrt{\Mbf}\right)^{-\top}f(\Sbf) \left(\sqrt{\Mbf}\right)^{-1}$
\item $\Amzero(\alp)\sim\mathcal{N}(\amzero,\alp)$ independent from $\Zbf.$
\end{itemize} 
It directly provides the covariance matrix of $\tilde{\Ybf}_\alp:$
$$\tilde{\sigbf}_\alp = \sigbf + \alp \phibf \phibf ^\top.$$

Now let us show that $\tilde{\sigbf}_\alp \tilde{\qbf}_\alp = I_{m},$ with $\tilde{\qbf}_\alp = \sqrt{\Mbf}\Sbf^2\sqrt{\Mbf}^\top + \frac{1}{\alp}\left(\Mbf\phibf\right)\left(\Mbf\phibf\right)^\top.$ The proof for $\tilde{\qbf}_\alp\tilde{\sigbf}_\alp = I_{m}$ is identical and thus not provided here.

$$
\tilde{\sigbf}_\alp \tilde{\qbf}_\alp=\underbrace{\sigbf \sqrt{\Mbf} \Sbf^2 \sqrt{\Mbf}^\top}_{U_1}
+ \underbrace{\phibf \phibf^\top (\Mbf\phibf)(\Mbf\phibf)^\top}_{U_2}+
 \underbrace{\alp \phibf \phibf^\top \sqrt{\Mbf} \Sbf^2 \sqrt{\Mbf}^\top}_{U_3}+
 \underbrace{\frac{1}{\alp} \sigbf (\Mbf\phibf)(\Mbf\phibf)^\top}_{U_4},
$$

and let us detail the computation of $U_1,U_2,U_3$ and $U_4.$
\begin{enumerate}
\item We have $U_1 = \left(\sqrt{\Mbf}\right)^{-\top}\Vbf\diag\left(0,1,\dots,1\right)\Vbf^\top\sqrt{\Mbf}^\top$ as:

\begin{align*}
\sigbf\sqrt{\Mbf}\Sbf^2\sqrt{\Mbf}^\top &= \left(\sqrt{\Mbf}\right)^{-\top}\Vbf f(\Sbf) \Vbf^\top \left(\sqrt{\Mbf}\right)^{-1}\sqrt{\Mbf}\Vbf\Sbf^2 \Vbf^\top\sqrt{\Mbf}^\top \\
&= \left(\sqrt{\Mbf}\right)^{-\top}\Vbf f(\Sbf) \Sbf^2 \Vbf^\top\sqrt{\Mbf}^\top\\
&= \left(\sqrt{\Mbf}\right)^{-\top}\Vbf\diag\left(0,1,\dots,1\right)\Vbf^\top\sqrt{\Mbf}^\top
\end{align*}

\item From Appendix~\ref{first_scenario_appendix}, $\left(\sqrt{\Mbf}\right)^\top\phibf \in \text{Ker}(\Sbf)$ and has unit norm. Then, $\left(\sqrt{\Mbf}\right)^\top\phibf = \vbf_1$, with $\vbf_1$ the first column of $\Vbf.$ With $\phibf^\top \Mbf \phibf = 1$, it comes $U_2 =  \phibf\phibf^\top\Mbf^\top = \left(\sqrt{\Mbf}\right)^{-\top}\vbf_1\vbf_1^\top\sqrt{\Mbf}^\top.$ 
\item $U_3 = 0$ as $\left(\sqrt{\Mbf}\right)^\top\phibf \in \text{Ker}(\Sbf)$ (see Appendix~\ref{first_scenario_appendix}).
\item $U_4 = 0$ as $\Mbf\phibf\in\text{Ker}(\sigbf)$ (see Appendix~\ref{first_scenario_appendix})
\end{enumerate}

Finally, $\tilde{\sigbf}_\alp \tilde{\qbf} = U_1+U_2 = \left(\sqrt{\Mbf}\right)^{-\top}\left(\Vbf\diag\left(0,1,\dots,1\right)\Vbf^\top+ \vbf_1\vbf_1^{\top}\right)\sqrt{\Mbf}^{\top} = I_{m}.$

\section{Proof of equation~\eqref{relation_augment}}\label{appendix_augmented}

This section provides a proof of equation~\eqref{relation_augment} for the \emph{second scenario}:

$$
\ubf^{\star}_\noise = \unoisealp(\ybf) + \Bigg(\frac{\phibf - \hh\left[\unoisealp\left(\Abf_n \phibf\right)\right]} {\left(\Mbf\phibf\right)^\top\ubf_{\noise, \alp}\left(\Abf_n \phibf\right)} - \phibf + \hh\left[\unoisealp\left(\Abf_n \phibf\right)\right]\Bigg) \left(\Mbf\phibf\right)^\top\ubf_{\noise, \alp}(\ybf).
$$

The proof is identical for the \emph{first scenario}, i.e. for $\noise=0.$ In the following, we denote $\Kbf_\noise = \Abf_n\sigbf\Abf_n^\top+\noise^2\Ibf_n.$ 

\subsection{Reformulation of $\unoisealp(\xbf)$.}

Let us first reformulate $\unoisealp(\xbf) = \tilde{\qbf}_\alp^{-1} \Abf_n^\top \left( \Abf_n \tilde{\qbf}_\alp^{-1} \Abf_n^\top +\noise^2 \Ibf_n \right)^{-1} \xbf$ for any $\xbf \in\mathbb{R}^{n}.$
\begin{align}
\unoisealp(\xbf)&= \left(\sigbf \Abf_n^T + \alp \phibf \phibf^\top  \Abf_n^T\right)\left(\Abf_n\sigbf\Abf_n^\top + \noise^2\Ibf_n+\alp \Abf_n \phibf \phibf^\top \Abf_n^T\right)^{-1}\xbf \notag\\
&= \left(\sigbf \Abf_n^T + \alp \phibf \phibf^\top  \Abf_n^T\right)\left(\Kbf_\noise^{-1} - \alp \Kbf_\noise^{-1} \Abf_n \phibf(1 + \alp \phibf^\top \Abf_n^\top\Kbf_\noise^{-1}\Abf_n\phibf)^{-1}\phibf^\top \Abf_n^\top \Kbf_\noise^{-1}\right)\xbf \label{wood_for}
\end{align} 
from Woodbury Formula. Using the fact that
$$\alp \phibf^\top \Abf_n^\top \Kbf_\noise^{-1} \Abf_n \phibf
(1 + \alp \phibf^\top \Abf_n^\top \Kbf_\noise^{-1} \Abf_n \phibf)^{-1}
=
1 -
(1 + \alp \phibf^\top \Abf_n^\top \Kbf_\noise^{-1} \Abf_n \phibf)^{-1},$$
we obtain
\begin{align*}
\unoisealp(\xbf)
&= \sigbf \Abf_n^T 
\Big(
    \Kbf_\noise^{-1}
    - \alp \Kbf_\noise^{-1} \Abf_n \phibf
      (1 + \alp \phibf^\top \Abf_n^\top \Kbf_\noise^{-1} \Abf_n \phibf)^{-1}
      \phibf^\top \Abf_n^\top \Kbf_\noise^{-1}
\Big)\xbf \\
&\quad
+ \alp \phibf \phibf^\top \Abf_n^T \Kbf_\noise^{-1}\ybf \\
&\quad
- \alp \phibf
\Big(
    1 -
    (1 + \alp \phibf^\top \Abf_n^\top \Kbf_\noise^{-1} \Abf_n \phibf)^{-1}
\Big)
\phibf^\top \Abf_n^\top \Kbf_\noise^{-1}\xbf,
\end{align*}
Hence,
\begin{align}
\unoisealp(\xbf)&= \sigbf \Abf_n^\top \Kbf_\noise^{-1}\xbf + \left(\phibf -\sigbf\Abf_n^\top\Kbf_\noise^{-1}\Abf_n\phibf\right) \left(\frac{1}{\alp} +  \phibf^\top \Abf_n^\top\Kbf_\noise^{-1}\Abf_n\phibf\right)^{-1}\phibf^\top \Abf_n^\top \Kbf_\noise^{-1}\xbf \notag \\
&= \amalp(\xbf)\phibf + \sigbf\Abf_n^\top\Kbf_\noise^{-1}\left(\xbf - \amalp(\xbf)\Abf_n\phibf\right)\label{eq_a_alp}
\end{align}
where $\amalp(\xbf) = \left(\frac{1}{\alp} + \phibf^\top \Abf_n^\top\Kbf_\noise^{-1}\Abf_n\phibf\right)^{-1}\phibf^\top \Abf_n^\top\Kbf_\noise^{-1} \xbf.$

\subsection{Computation of $\hh\left[\unoisealp\left(\Abf_n \phibf\right)\right].$}

By definition, $$\hh\left[\unoisealp\left(\Abf_n \phibf\right)\right] = \frac{\unoisealp\left(\Abf_n \phibf\right) - \phibf\left(\Mbf\phibf\right)^\top \unoisealp\left(\Abf_n \phibf\right)}{1 - \left(\Mbf\phibf\right)^\top \unoisealp\left(\Abf_n \phibf\right)}.$$
Using Equation~\ref{eq_a_alp} together with $\Mbf\phibf \in \ker(\sigbf),$ we obtain $$\forall \xbf \in\mathbb{R}^n, ~~\left(\Mbf\phibf\right)^\top\ubf_{\noise, \alp}(\xbf) = \amalp(\xbf)\phibf^\top\Mbf\phibf = \amalp(\xbf).$$
Finally, we have
\begin{align*}
\hh\left[\unoisealp\left(\Abf_n \phibf\right)\right] &= \frac{\unoisealp\left(\Abf_n \phibf\right) - \amalp\left(\Abf_n \phibf\right)\phibf}{1-\amalp\left(\Abf_n \phibf\right)} \\
&=  \sigbf\Abf_n^\top\Kbf_\noise^{-1}\Abf_n \phibf \text{ from equation~\eqref{eq_a_alp}}.
\end{align*}

\subsection{Full derivation}

\begin{align*}
\frac{\phibf - \hh\left[\unoisealp\left(\Abf_n \phibf\right)\right]} {\left(\Mbf\phibf\right)^\top\ubf_{\noise, \alp}\left(\Abf_n \phibf\right)}\left(\Mbf\phibf\right)^\top\ubf_{\noise, \alp}(\ybf) &= \left(\phibf - \hh\left[\unoisealp\left(\Abf_n \phibf\right)\right]\right)\frac{\amalp(\ybf)}{\amalp(\Abf_n\phibf)} \\
&= \left(\phibf - \hh\left[\unoisealp\left(\Abf_n \phibf\right)\right]\right)\frac{\phibf^\top \Abf_n^\top\Kbf_\noise^{-1}\ybf}{\phibf^\top \Abf_n^\top\Kbf_\noise^{-1} \Abf_n\phibf} \\
&= \amnoise\left(\phibf - \hh\left[\unoisealp\left(\Abf_n \phibf\right)\right]\right)
\end{align*}
by definition of $\amnoise$ (see Equation~\ref{eq_amnoise}). Finally, combining the previous results yields
\begin{align*}
&\unoisealp(\ybf) +  \Bigg(\frac{\phibf - \hh\left[\unoisealp\left(\Abf_n \phibf\right)\right]} {\left(\Mbf\phibf\right)^\top\ubf_{\noise, \alp}\left(\Abf_n \phibf\right)} - \phibf + \hh\left[\unoisealp\left(\Abf_n \phibf\right)\right]\Bigg) \left(\Mbf\phibf\right)^\top\ubf_{\noise, \alp}(\ybf) \\
=~& \amalp(\ybf)\phibf + \sigbf\Abf_n^\top\Kbf_\noise^{-1}\left(\ybf - \amalp(\ybf)\Abf_n\phibf\right) + \left(\amnoise- \amalp(\ybf)\right)\left(\phibf - \hh\left[\unoisealp\left(\Abf_n \phibf\right)\right]\right)\\
=~&\amalp(\ybf)\phibf + \sigbf\Abf_n^\top\Kbf_\noise^{-1}\left(\ybf - \amalp(\ybf)\Abf_n\phibf\right) + \left(\amnoise- \amalp(\ybf)\right)\left(\phibf -  \sigbf\Abf_n^\top\Kbf_\noise^{-1}\Abf_n \phibf\right)\\
=~& \amnoise\left(\phibf -\sigbf\Abf_n^\top\Kbf_\noise^{-1}\Abf_n \phibf\right) + \sigbf\Abf_n^\top\Kbf_\noise^{-1}\ybf \\
=~&\ubf^{\star}_\noise.
\end{align*}

\section{Cholesky decomposition for sparse linear systems}\label{cholesky_appendix}

This section details the computation of the matrix $ \Bbf^{-1} \Nbf $, where $ \Bbf $ is a large symmetric matrix of size $ k_1 \times k_1 $ with Cholesky decomposition $ \Bbf = \Lbf \Lbf^\top $, and $ \Nbf $ is a matrix of size $ k_1 \times k_2 $.

\noindent The idea is to solve the following two triangular linear systems:
$$
\Lbf \ombf_1 = \Nbf \quad \text{and} \quad \Lbf^\top \ombf_2 = \ombf_1,
$$
\noindent in order to compute $ \ombf_2 = \Bbf^{-1} \Nbf $.

\section{Woodbury formula for the second scenario}\label{appendix_woodbury}

This section proves equation~\eqref{eq_noise_augment}:

$$\tilde{\qbf}_\alp^{-1} \Abf_n^\top\left( \Abf_n \tilde{\qbf}_\alp^{-1} \Abf_n^\top + \noise^2\Ibf_n\right)^{-1}\xbf = \left(\noise^2\tilde{\qbf}_\alp+\Abf_n^\top\Abf_n\right)^{-1}\Abf_n^\top\xbf,$$

for $\noise > 0.$ We have 
\begin{align*}
&\tilde{\qbf}_\alp^{-1} \Abf_n^\top\left( \Abf_n \tilde{\qbf}_\alp^{-1} \Abf_n^\top + \noise^2\Ibf_n\right)^{-1} \\&= \tilde{\qbf}_\alp^{-1}\Abf_n^\top \left[\noise^{-2}\Ibf_n - \noise^{-2}\Abf_n\left(\noise^2\tilde{\qbf}_\alp + \Abf_n^\top\Abf_n\right)^{-1}\Abf_n^\top\right] \text{ from Woodbury formula}\\
&= \left(\noise^2\tilde{\qbf}_\alp+\Abf_n^\top\Abf_n\right)^{-1}\Abf_n^\top - \left(\Ibf_m + \noise^{-2}\tilde{\qbf}_\alp^{-1}\Abf_n^\top\Abf_n\right)\left(\noise^2\tilde{\qbf}_\alp+\Abf_n^\top\Abf_n\right)^{-1}\Abf_n^\top + \noise^{-2}\tilde{\qbf}_\alp^{-1}\Abf_n^\top\\
& = \left(\noise^2\tilde{\qbf}_\alp+\Abf_n^\top\Abf_n\right)^{-1}\Abf_n^\top - \noise^{-2}\tilde{\qbf}_\alp^{-1}\Abf_n^\top+\noise^{-2}\tilde{\qbf}_\alp^{-1}\Abf_n^\top\\
& = \left(\noise^2\tilde{\qbf}_\alp+\Abf_n^\top\Abf_n\right)^{-1}\Abf_n^\top.
\end{align*}

\section{Selection of the parameter $\alpha$}\label{appendix_alpha}

In Section~\ref{sec_alp}, it is stated that the parameter $\alpha$ must satisfy
\[
\frac{1}{\sqrt{\alpha}} \in \big[\lambda_{1}^{(m)},\, \lambda_{m-1}^{(m)}\big],
\]
where $\lambda_{0}^{(m)} \le \lambda_{1}^{(m)} \le \cdots \le \lambda_{m-1}^{(m)}$ denote the eigenvalues of $\Sbf$. We now describe a procedure to determine such a value of $\alpha$ by computing $\left(\lambda_{1}^{(m)}\right)^2$ and $\left(\lambda_{m-1}^{(m)}\right)^2$.

The largest eigenvalue of a matrix can be computed, for instance, using the power iteration method~\cite{mises1929praktische}, which only requires matrix--vector products. Algorithm~\ref{power_iteration} details this method for the matrix $\Sbf^2$, providing an estimate of $\left(\lambda_{m-1}^{(m)}\right)^2$.

\begin{algorithm}[H]
\caption{Power Iteration Method for $\left(\lambda_{m-1}^{(m)}\right)^2.$}
\begin{algorithmic}[1]
\Require Matrix $\Sbf^2 \in \mathbb{R}^{m \times m}$, initial vector $\bbf_0 \in \mathbb{R}^m$, tolerance $\epsilon > 0$, max iterations $N$
\State $\bbf \gets \frac{\bbf_0}{\|\bbf_0\|_2}$ 
\State $err \gets \infty, \quad i \gets 0$
\While{$err > \epsilon$ and $i < N$}
    \State $\wbf \gets \Sbf^2 \bbf$
    \State $\lambda \gets \bbf^T \wbf$ 
    \State $err \gets \|\wbf - \lambda \bbf\|_2$ 
    \State $\bbf \gets \frac{\wbf}{\|\wbf\|_2}$
    \State $i \gets i+1$
\EndWhile
\State \Return $(\lambda, err)$ the highest eigenvalue of $\Sbf^2$ and the residual
\end{algorithmic}\label{power_iteration}
\end{algorithm}

Moreover, $\left(\lambda_{1}^{(m)}\right)^2$ is the largest eigenvalue of
\[
\left(\Sbf^2+\Ibf_{m}\right)^{-1} - \bigl(\sqrt{\Mbf}^\top \phibf\bigr)\bigl(\sqrt{\Mbf}^\top \phibf\bigr)^\top,
\]
and can be computed using the power iteration method together with the Cholesky decomposition of $\Sbf^2 + \Ibf_m$, as detailed in Algorithm~\ref{power_iteration_2}.

\begin{algorithm}[H]
\caption{Power Iteration Method for the second eigenvalue for $\left(\lambda_{1}^{(m)}\right)^2.$}
\begin{algorithmic}[1]
\Require Matrix $\Sbf^2 \in \mathbb{R}^{m \times m}$, first eigenvector $\vbf = \sqrt{\Mbf}^\top\phibf \in \mathbb{R}^m$, initial vector $\bbf_0 \in \mathbb{R}^m$, tolerance $\epsilon > 0$, max iterations $N$
\State Compute Cholesky decomposition of $\Sbf^2 + \Ibf_m$
\State $\bbf \gets \bbf_0 / \|\bbf_0\|_2$ 
\State $err \gets \infty, \quad i \gets 0$
\While{$err > \epsilon$ \textbf{and} $i < N$}
    \State Solve $(\Sbf^2 + \Ibf_m)\wbf = \bbf$ using Cholesky decomposition
    \State $\wbf \gets \wbf - \vbf(\vbf^\top \bbf)$ 
    \State $\lambda \gets \bbf^\top \wbf$ 
    \State $err \gets \|\wbf - \lambda \bbf\|_2$ 
    \State $\bbf \gets \wbf / \|\wbf\|_2$
    \State $i \gets i + 1$
\EndWhile
\State $\lambda \gets \frac{1}{\lambda} - 1$
\State $err \gets \|\Sbf^2\bbf - \lambda\bbf\|_2$
\State \Return $(\lambda, err)$ the estimated second eigenvalue of $\Sbf^2$ and the residual
\end{algorithmic}\label{power_iteration_2}
\end{algorithm}

The final residuals computed in Algorithms~\ref{power_iteration} and~\ref{power_iteration_2} allow one to control the error in the estimation of $\lambda_{1}^{(m)}$ and $\lambda_{m-1}^{(m)}$, thanks to Proposition~\ref{error_eigen}.

\begin{prop}[Error Bound on the Eigenvalue]\label{error_eigen}
Let $\bbf \in \mathbb{R}^{m}$ be a unit vector such that $\|\bbf\|_2 = 1$. Let $\lambda \in \mathbb{R}$ and $\epsilon > 0$.

If the residual satisfies:
$$\|\Sbf^2\bbf - \lambda\bbf\|_2 < \epsilon$$
Then there exists an exact eigenvalue $\mu \in \text{spec}(\Sbf^2)$ such that:
$$\lvert \mu - \lambda \rvert < \epsilon$$
\end{prop}

\begin{proof}
Let us define the residual vector $\rbf=\Sbf^2\bbf - \lambda\bbf.$ It follows that $$\left(\Sbf^2-\rbf\bbf^\top\right)\bbf = \lambda\bbf,$$ implying that $\lambda \in \text{spec}(\Sbf^2-\rbf\bbf^\top).$ 

Since $\Sbf^2$ is symmetric, it is diagonalizable by an orthogonal matrix, meaning its spectral condition number is $1$. By the Bauer-Fike theorem~\cite{bauer1960norms}, there exists $\mu \in \text{spec}(\Sbf^2)$ such that: $$\lvert \mu - \lambda \rvert \leq \| \rbf\bbf^\top \|_2 = \|\rbf\|_2 < \epsilon$$
\end{proof}

Then, to ensure that $\frac{1}{\alpha} \in \big[\left(\lambda_{1}^{(m)}\right)^2,\, \left(\lambda_{m-1}^{(m)}\right)^2\big]$, the procedure described in Algorithm~\ref{algo_alpha} is applied. In practice, $\lambda_{m-1}^{(m)}$ is typically several orders of magnitude larger than $\lambda_{1}^{(m)}$. Consequently, the maximum number of iterations $N$ remains moderate, and a conservative margin coefficient $\gamma$ can be employed to ensure spectral separation. In our numerical experiments, we set $N = 1000$ and $\gamma = 100$.

\begin{algorithm}[H]
\caption{Overall procedure to select a value for $\alpha$}
\begin{algorithmic}[1]
\Require Matrix $\Sbf^2 \in \mathbb{R}^{m \times m}$, first eigenvector $\vbf = \sqrt{\Mbf}^\top\phibf \in \mathbb{R}^m$, initial vector $\bbf_0 \in \mathbb{R}^m$, max iterations $N$, margin coefficient $\gamma$
\State \label{step:comp1} Compute estimate $\tilde{\lambda}_1$ of $(\lambda_{1}^{(m)})^2$ using Algorithm~\ref{power_iteration_2} with $N$ iterations; store residual $err_1$
\State Compute estimate $\tilde{\lambda}_{m-1}$ of $(\lambda_{m-1}^{(m)})^2$ using Algorithm~\ref{power_iteration} with $N$ iterations; store residual $err_{m-1}$
\If{$\tilde{\lambda}_1 + err_1 \geq \tilde{\lambda}_{m-1} - err_{m-1}$}
\State $N \gets 2N$
\State \textbf{goto} Step~\ref{step:comp1}
\ElsIf{$\tilde{\lambda}_1 + \gamma \cdot err_1 > \tilde{\lambda}_{m-1} - err_{m-1}$}
    \State $\gamma \gets \max\left(\gamma / 2,1\right)$ 
    \State \textbf{goto} Step~\ref{step:comp1} 
\EndIf
\State $\alpha \gets \frac{1}{\tilde{\lambda}_1 + \gamma \cdot err_1}$
\State \Return $\alpha$
\end{algorithmic}\label{algo_alpha}
\end{algorithm}

\section{Likelihood estimation}\label{likeli}

This section details the computation of 

$$\llik_\noise(\param) =  -\frac{1}{2} \left(n\log(2\pi)+\log \lvert \Kbf_\noise \rvert + \left(\ybf - \amnoise \Abf_n \phibf\right)^\top \Kbf_\noise^{-1}\left(\ybf - \amnoise \Abf_n \phibf\right) \right).$$

The corresponding steps are summarized in Algorithm~\ref{ll_scenar1} and Algorithm~\ref{ll_scenar2}, for the first scenario and the second scenario, respectively.

Here, all quantities $\Kbf_\noise, \amnoise, \amalp, \Mbf, \unoisealp, \phibf$ depend on $\param$, but for simplicity, this dependence is not explicitly indicated.

\subsection{Compute $\ell^{(1)}_\noise(\param) = \left(\ybf - \amnoise \Abf_n \phibf\right)^\top \Kbf_\noise^{-1}\left(\ybf - \amnoise \Abf_n \phibf\right)$}

\subsubsection{Tractable formula for $\ell^{(1)}_\noise(\param)$ }

We have 
\begin{align*}
\ell^{(1)}_\noise(\param) &= \left(\ybf - \amnoise \Abf_n \phibf\right)^\top\Kbf_\noise^{-1}\left(\ybf - \amnoise \Abf_n \phibf\right) \\
& = \ybf^\top \Kbf_\noise^{-1}\ybf - 2\amnoise\phibf^\top\Abf_n^\top \Kbf_\noise^{-1}\ybf + \left(\amnoise\right)^{2} \Abf_n^\top \Kbf_\noise^{-1} \Abf_n\phibf \\
& = \ybf^\top \Kbf_\noise^{-1}\ybf - \amnoise\phibf^\top\Abf_n^\top \Kbf_\noise^{-1}\ybf \text{ from the definition of }\amnoise \text{(see equation~\eqref{eq_amnoise}),}
\end{align*}
and we will show that 

\begin{equation}\label{eq_l1}
\ell^{(1)}_\noise(\param) = \ybf^\top \left( \Abf_n \tilde{\qbf}_\alp^{-1} \Abf_n^\top + \noise^2\Ibf_n\right)^{-1}\ybf - \frac{1}{\alp}\frac{\amalp(\ybf)^2}{\amalp(\Abf_n\phibf)}.
\end{equation}

From equation~\eqref{wood_for} of Appendix~\ref{appendix_augmented}, we have 

$$\ybf^\top \left( \Abf_n \tilde{\qbf}_\alp^{-1} \Abf_n^\top + \noise^2\Ibf_n\right)^{-1}\ybf = \ybf^\top \Kbf_\noise^{-1}\ybf - \amalp(\ybf)\ybf^\top\Kbf_\noise^{-1}\Abf_n\phibf.$$

Then, 
\begin{align*}
&\ybf^\top \left( \Abf_n \tilde{\qbf}_\alp^{-1} \Abf_n^\top + \noise^2\Ibf_n\right)^{-1}\ybf - \frac{1}{\alp}\frac{\amalp(\ybf)^2}{\amalp(\Abf_n\phibf)}\\ 
&= \ybf^\top \Kbf_\noise^{-1}\ybf - \amalp(\ybf)\ybf^\top\Kbf_\noise^{-1}\Abf_n\phibf\left(1 + \frac{1}{\alp\phibf^\top\Abf_n^\top \Kbf_\noise^{-1}\Abf_n\phibf}\right)\\
&= \ybf^\top \Kbf_\noise^{-1}\ybf - \amalp(\ybf)\ybf^\top\Kbf_\noise^{-1}\Abf_n\phibf\frac{\phibf^\top\Abf_n^\top \Kbf_\noise^{-1}\Abf_n\phibf+\frac{1}{\alp}}{\phibf^\top\Abf_n^\top \Kbf_\noise^{-1}\Abf_n\phibf}\\ 
&= \ybf^\top \Kbf_\noise^{-1}\ybf - \amnoise\phibf^\top\Abf_n^\top \Kbf_\noise^{-1}\ybf \\
&= \ell^{(1)}_\noise(\param).
\end{align*}

\subsubsection{First scenario}

Here, $\noise = 0,$ $\Abf_n  \tilde{\qbf}_\alp^{-1}\Abf_n^\top = \left[\tilde{\qbf}_\alpha^{-1}\right]_{I,I},$ and we compute $\ell^{(1)}_0(\param)$ using equation~\eqref{eq_l1}.

\begin{itemize}
\item $\left( \Abf_n \tilde{\qbf}_\alpha^{-1} \Abf_n^\top \right)^{-1}$ corresponds to a Schur complement and is given by:
$$
\left( \Abf_n \tilde{\qbf}_\alpha^{-1} \Abf_n^\top \right)^{-1} = \left[\tilde{\qbf}_\alpha\right]_{I,I} - \left[\tilde{\qbf}_\alpha\right]_{I,\bar{I}}\left[\tilde{\qbf}_\alpha\right]_{\bar{I},\bar{I}}^{-1}\left[\tilde{\qbf}_\alpha\right]_{\bar{I},I}.
$$

Then, $\ybf^\top\left( \Abf_n \tilde{\qbf}_\alpha^{-1} \Abf_n^\top \right)^{-1}\ybf=  \ybf^\top\left[\tilde{\qbf}_\alpha\right]_{I,I}\ybf^\top - \ybf^\top\left[\tilde{\qbf}_\alpha\right]_{I,\bar{I}}\left[\tilde{\qbf}_\alpha\right]_{\bar{I},\bar{I}}^{-1}\left[\tilde{\qbf}_\alpha\right]_{\bar{I},I}\ybf,$ and
can be computed using the Cholesky decomposition of $\left[\tilde{\qbf}_\alpha\right]_{\bar{I},\bar{I}},$ as described in Section~\ref{first_scenar}.
\item From Appendix~\ref{appendix_augmented}, we have for $\xbf \in \mathbb{R}^n,$

\begin{align*}
\amalpzero(\xbf) &= \left(\Mbf\phibf \right)^\top\ubf_{0,\alp}(\xbf) \\
&=\left(\Mbf\phibf \right)^\top\tilde{\qbf}_\alp^{-1} \Abf_n^\top \left( \Abf_n \tilde{\qbf}_\alp^{-1} \Abf_n^\top \right)^{-1}\xbf \\
&=\left(\Mbf\phibf \right)^\top\left(\sigbf + \alp \phibf \phibf^\top\right)\Abf_n^\top \left( \Abf_n \tilde{\qbf}_\alp^{-1} \Abf_n^\top\right)^{-1}\xbf\\
&=\alp\phibf^\top\Abf_n^\top \left( \Abf_n \tilde{\qbf}_\alp^{-1} \Abf_n^\top \right)^{-1}\xbf \text{ using } \phibf^\top\Mbf\phibf = 1,
\end{align*}
and can be computed again using Schur complement and Cholesky decomposition.
\end{itemize}

\subsubsection{Second scenario}

Here, $\noise > 0$ and we compute $\ell^{(1)}_\noise(\param)$ using equation~\eqref{eq_l1}.

\begin{itemize}
\item From Woodbury formula, we have

\begin{align*}
\ybf^\top\left( \Abf_n \tilde{\qbf}_\alp^{-1} \Abf_n^\top + \noise^2\Ibf_n\right)^{-1}\ybf &= \ybf^\top\left[\noise^{-2}\Ibf_n - \noise^{-2}\Abf_n\left(\noise^2\tilde{\qbf}_\alp + \Abf_n^\top\Abf_n\right)^{-1}\Abf_n^\top\right]\ybf \\
& =\tau^{-2}\left[\ybf^\top\ybf-\ybf^\top\Abf_n\unoisealp(\ybf)\right]
\end{align*} which is tractable (see Section~\ref{second_scenar}).
\item From Appendix~\ref{appendix_augmented}, we have for $\xbf \in \mathbb{R}^n,$

$$\amalp(\xbf) = \left(\Mbf\phibf \right)^\top\ubf_{\noise,\alp}(\xbf).$$
\end{itemize}

\subsection{Compute $\ell^{(2)}_\noise(\param) = \log(\Kbf_\noise)$}

We have 
\begin{align*}
\ell^{(2)}_\noise(\param) &= \log \lvert \Abf_n\sigbf\Abf_n^\top+\noise^2\Ibf_n\rvert \\
&=\log \lvert \Abf_n \tilde{\qbf}_\alp^{-1}\Abf_n^\top +\noise^2\Ibf_n \rvert + \log \lvert 1 - \alp (\Abf_n \phibf)^\top \left(\Abf_n \tilde{\qbf}_\alp^{-1}\Abf_n^\top+\noise^2\Ibf_n\right)^{-1} \Abf_n \phibf \rvert\\
&=\log \lvert \Abf_n \tilde{\qbf}_\alp^{-1}\Abf_n^\top +\noise^2\Ibf_n \rvert + \log \lvert 1 - \amalp(\Abf_n\phibf) \rvert
\end{align*}

\subsubsection{First scenario}

Here, $\noise = 0$ and $\Abf_n  \tilde{\qbf}_\alp\Abf_n^\top = \left[\tilde{\qbf}_\alpha^{-1}\right]_{I,I}.$ Then, 

\begin{itemize}
\item $\lvert \Abf_n \tilde{\qbf}_\alp^{-1}\Abf_n^\top\rvert = \frac{\lvert \tilde{\qbf}_\alpha \rvert}{\lvert \left[\tilde{\qbf}_\alpha\right]_{\bar{I},\bar{I}} \rvert }$, that can be computed using the Cholesky decompositions of $\tilde{\qbf}_\alpha$ and $\left[\tilde{\qbf}_\alpha\right]_{\bar{I},\bar{I}}.$
\item $\amalpzero(\Abf_n\phibf)$ is already computed for $\ell^{(1)}_0(\param)$.
\end{itemize}

\subsubsection{Second scenario}

Here, $\noise > 0,$ and then

\begin{itemize}
\item $
\log\lvert \Abf_n\tilde{\qbf}_\alpha^{-1} \Abf_n^\top+\noise^2\Ibf_n\rvert = n\log(\noise^2)+\log\lvert \Ibf_m + \noise^{-2}\tilde{\qbf}_\alpha^{-1}\Abf_n^\top\Abf_n\rvert 
= (n-m)\log(\noise^2)-\log\lvert\tilde{\qbf}_\alpha\rvert +\log\lvert \noise^2\tilde{\qbf}_\alpha+\Abf_n^\top\Abf_n\rvert
$, that can be computed using the Cholesky decompositions of $\noise^2\tilde{\qbf}_\alpha+\Abf_n^\top\Abf_n$ and $\tilde{\qbf}_\alpha.$
\item $\amalp(\Abf_n\phibf)$ is already computed for $\ell^{(1)}_\noise(\param)$.
\end{itemize}

\begin{algorithm}[H]
\caption{Computation of the log-likelihood in the first scenario}
\begin{algorithmic}[1]
\Require $\Mbf, \Fbf, \phibf, I, \alpha, \Abf_n$
\State $\Sbf \gets \sqrt{\Mbf}^{-1} \, \Fbf \, \sqrt{\Mbf}^{-\top}$
\State $\qbf \gets \sqrt{\Mbf} \, \Sbf^2 \, \sqrt{\Mbf}^{\top}$
\State Compute Cholesky of $\left[\tilde{\qbf}_\alpha\right]_{\bar{I},\bar{I}} = \left[\qbf\right]_{\bar{I},\bar{I}} + \frac{1}{\alpha}\left[\Mbf\phibf\right]_{\bar{I}}\left[\Mbf\phibf\right]_{\bar{I}}^\top$ using sparsity of $\left[\tilde{\qbf}_\alpha\right]_{\bar{I},\bar{I}}$ and a rank-one update
\State Solve $\left[\tilde{\qbf}_\alpha\right]_{\bar{I},\bar{I}}\ombf_\ybf = \left[\tilde{\qbf}_\alpha\right]_{\bar{I},I}\ybf$
\State $\vbf(\ybf) \gets \left[\qbf\right]_{I,I} \ybf - \left[\qbf\right]_{I,\bar{I}}\ombf_\ybf + \frac{1}{\alpha}\left[\Mbf\phibf\right]_{I}\left[\Mbf\phibf\right]_{I}^\top\ybf -  \frac{1}{\alpha}\left[\Mbf\phibf\right]_{I}\left[\Mbf\phibf\right]_{\bar{I}}^\top\ombf_\ybf$
\State $\amalp(\ybf) \gets \alp\phibf^\top \Abf_n^\top \vbf(\ybf)$
\State Solve $\left[\tilde{\qbf}_\alpha\right]_{\bar{I},\bar{I}}\ombf_{\Abf_n\phibf} = \left[\tilde{\qbf}_\alpha\right]_{\bar{I},I}\Abf_n\phibf$
\State $\vbf(\Abf_n\phibf) \gets \left[\qbf\right]_{I,I} \Abf_n\phibf - \left[\qbf\right]_{I,\bar{I}}\ombf_{\Abf_n\phibf} + \frac{1}{\alpha}\left[\Mbf\phibf\right]_{I}\left[\Mbf\phibf\right]_{I}^\top\Abf_n\phibf -  \frac{1}{\alpha}\left[\Mbf\phibf\right]_{I}\left[\Mbf\phibf\right]_{\bar{I}}^\top\ombf_{\Abf_n\phibf}$
\State $\amalp(\Abf_n\phibf) \gets \alp\phibf^\top \Abf_n^\top \vbf(\Abf_n\phibf)$
\State $\ell^{(1)}_0 \gets \ybf^\top \vbf(\ybf) - \frac{1}{\alp}\frac{\amalp(\ybf)^2}{\amalp(\Abf_n\phibf)}$
\State Compute Cholesky of $\tilde{\qbf}_\alpha = \qbf + \frac{1}{\alpha}(\Mbf\phibf)(\Mbf\phibf)^\top$ using sparsity of $\qbf$ and a rank-one update
\State $\ell^{(2)}_0 \gets \log\lvert \tilde{\qbf}_\alpha \rvert -  \log\lvert\left[\tilde{\qbf}_\alpha\right]_{\bar{I},\bar{I}} \rvert - \log\lvert 1 - \amalp(\Abf_n\phibf)\rvert$
\State \Return $\llik_\noise(\param) = -\frac{1}{2}\left(n\log(2\pi)+\ell^{(1)}_0+\ell^{(2)}_0\right)$
\end{algorithmic}
\footnotesize\textit{Note:} 
$\vbf(\ybf) = \left(\Abf_n \tilde{\qbf}_\alpha^{-1} \Abf_n^\top\right)^{-1}\ybf$ \; and \; 
$\vbf(\Abf_n\phibf) = \left(\Abf_n \tilde{\qbf}_\alpha^{-1} \Abf_n^\top\right)^{-1}\Abf_n\phibf$.\label{ll_scenar1}
\end{algorithm}

\begin{algorithm}[H]
\caption{Computation of the log-likelihood in the second scenario}
\begin{algorithmic}[1]
\Require $\Mbf, \Fbf, \phibf, I, \alpha, \Abf_n$
\State $\Sbf \gets \sqrt{\Mbf}^{-1} \, \Fbf \, \sqrt{\Mbf}^{-\top}$
\State $\qbf \gets \sqrt{\Mbf} \, \Sbf^2 \, \sqrt{\Mbf}^{\top}$
\State Compute Cholesky of $\noise^2\tilde{\qbf}_\alp + \Abf_n^\top\Abf_n = \noise^2\qbf + \Abf_n^\top\Abf_n + \frac{\noise^2}{\alp}\left(\Mbf\phibf\right)\left(\Mbf\phibf\right)^\top$ using sparsity of $\noise^2\qbf+ \Abf_n^\top\Abf_n$ and a rank-one update
\State Solve $\left(\noise^2\tilde{\qbf}_\alp + \Abf_n^\top\Abf_n\right)\ombf_\ybf = \Abf_n^\top\ybf$
\State Solve $\left(\noise^2\tilde{\qbf}_\alp + \Abf_n^\top\Abf_n\right)\ombf_{\Abf_n\phibf} = \Abf_n^\top\Abf_n\phibf$
\State $\amalp(\ybf) \gets \left(\Mbf\phibf \right)^\top\ombf_\ybf$
\State $\amalp(\Abf_n\phibf) \gets \left(\Mbf\phibf \right)^\top\ombf_{\Abf_n\phibf}$
\State $\ell^{(1)}_\noise \gets \noise^{-2}\left(\ybf^\top\ybf-\ybf^\top\Abf_n\ombf_\ybf\right)-\frac{1}{\alp}\frac{\amalp(\ybf)^2}{\amalp(\Abf_n\phibf)}$
\State Compute Cholesky of $\tilde{\qbf}_\alpha = \qbf + \frac{1}{\alpha}(\Mbf\phibf)(\Mbf\phibf)^\top$ using sparsity of $\qbf$ and a rank-one update
\State $\ell^{(2)}_\noise \gets (n-m)\log(\noise^2)-\log\lvert\tilde{\qbf}_\alp \rvert + \log\lvert\noise^2\tilde{\qbf}_\alp + \Abf_n^\top\Abf_n\rvert + \log\lvert 1 - \amalp(\Abf_n\phibf)\rvert$
\State \Return $\llik_\noise(\param) = -\frac{1}{2}\left(n\log(2\pi)+\ell^{(1)}_\noise+\ell^{(2)}_\noise\right)$
\end{algorithmic}\label{ll_scenar2}
\end{algorithm}

\section{2D visualization of the results and errors}\label{appendix_2d}

Here, the results are presented in two dimensions using both spherical and cylindrical coordinates. 

\subsection{Analytical function on the sphere}

\begin{figure}[h]
\centering
\includegraphics[width=0.7\textwidth]{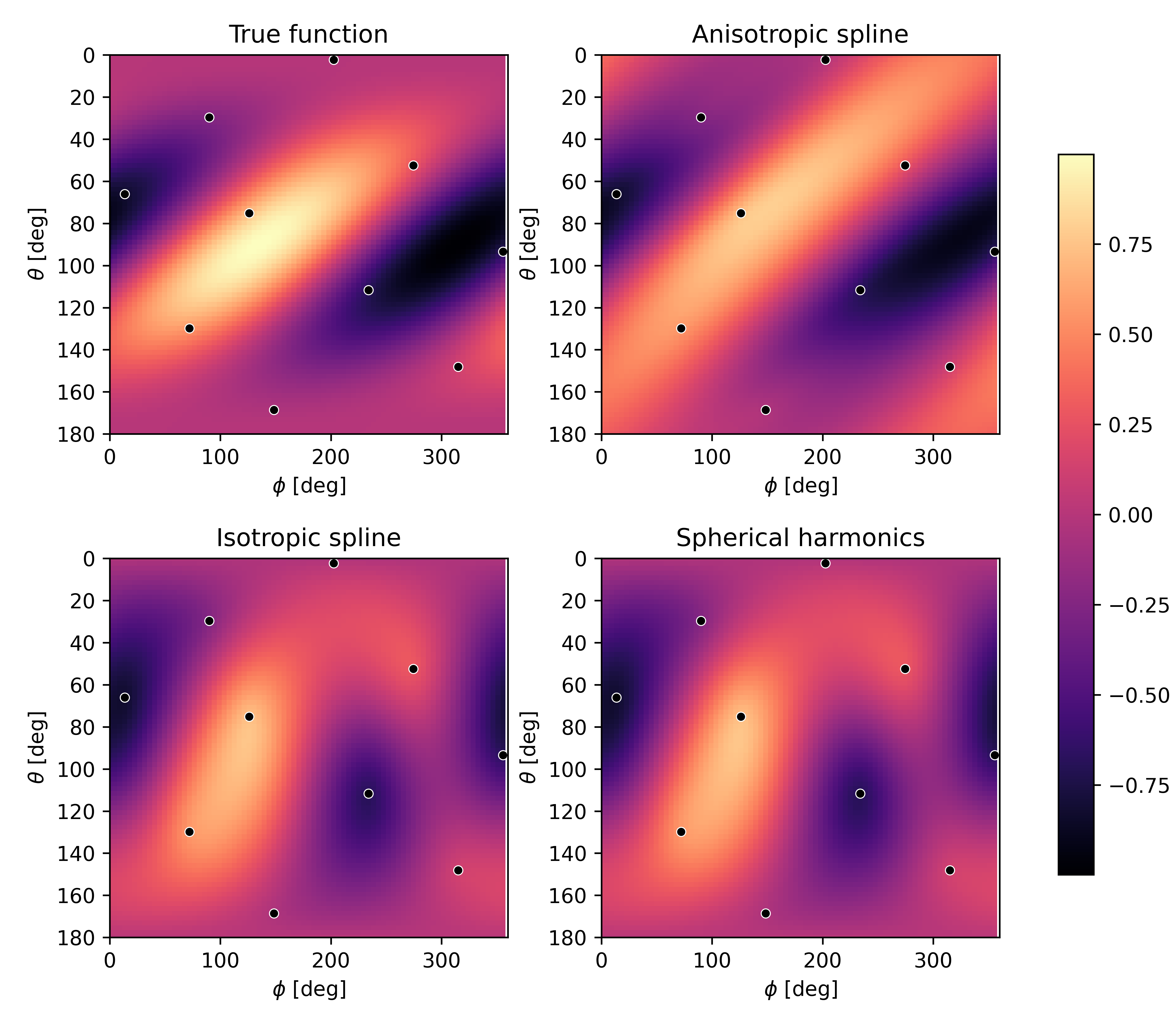}
\caption{Results for the analytical function on the sphere in the first scenario illustrated in 2D using the spherical coordinates $(\theta,\phi).$ $n=10$ observation points are shown as black dots. 
Top left: true function. 
Top right: prediction with anisotropies induced in the local charts defined by $(\theta, \phi)$. 
Bottom left: isotropic splines (no space deformation). 
Bottom right: splines using a kernel based on spherical harmonics.}
\label{results_sphere_2D}
\end{figure}

\FloatBarrier

\subsection{Real-world data on the sphere}

\subsubsection{Validation and anisotropies}

\begin{figure}[h]
\centering
\includegraphics[width=0.7\textwidth]{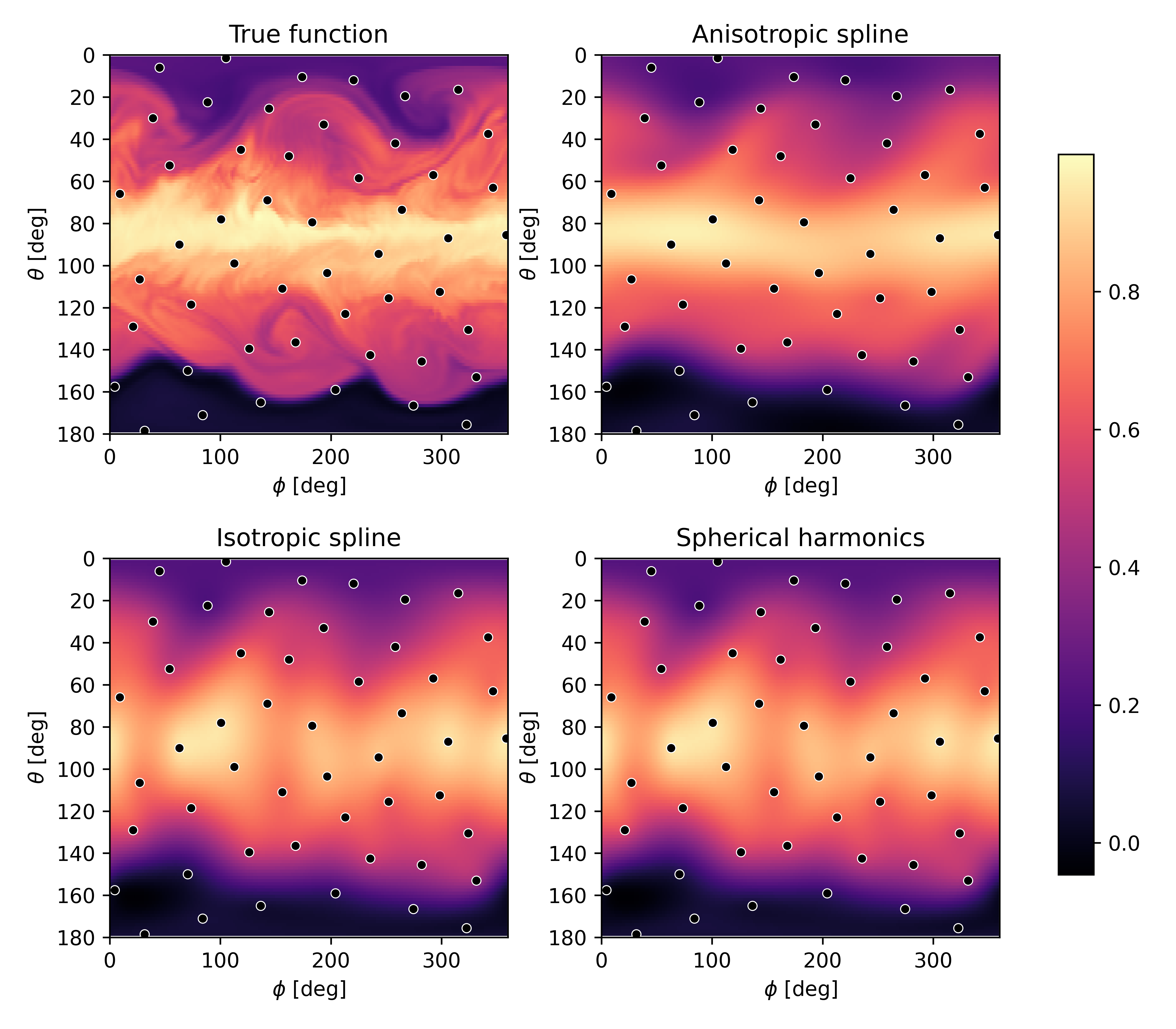}
\caption{Results for the normalized concentration of carbon dioxide on the sphere in the first scenario illustrated in 2D using the spherical coordinates $(\theta,\phi).$ $n=50$ observation points are shown as black dots. 
Top left: true function. 
Top right: prediction with anisotropies induced in the local charts defined by $(\theta, \phi)$. 
Bottom left: isotropic splines (no space deformation). 
Bottom right: splines using a kernel based on spherical harmonics.}
\label{results_earth_2D}
\end{figure}
\FloatBarrier

\subsubsection{Comparison with Mat\'ern solution}\label{matern}

To compare the spline prediction, which uses a random field $Z$ of covariance kernel $K_1,$ with a standard kriging model, we work with an a priori random field $\tilde{Z}$ as the solution to the Stochastic Partial Differential Equation (SPDE)

$$\left(\kappa^2 - \Delta\right)\tilde{Z} = W,$$
where $W$ is a Gaussian white noise on $\manif$ and $\kappa^2 > 0.$ In $\mathbb{R}^d,$ these solutions are known to exhibit the Mat\'ern covariance kernel. Following the approach of \cite{pereira_desassis_allard_2022}, the results are analogous to those presented here but include the scaling parameter appearing in the covariance kernel of the solution

$$\tilde{K}(\sbf_1,\sbf_2) = \sum_{k\in\mathbb{N}} \tilde{f}(\lambda_k) \phi_k(\sbf_1)\phi_k(\sbf_2),$$
with $\tilde{f}(\lambda) = \frac{1}{\left(\kappa^2+\lambda\right)^2}.$

The finite-element solution then yields a GMRF with a precision matrix defined as 

$$\tilde{Q}_\text{matern} = \sqrt{\Mbf} \left(\kappa^2\Ibf_m + \Sbf\right)^2 \sqrt{\Mbf}^{\top},$$

where classical posterior computations can be conducted because this matrix is sparse and invertible due to the scaling parameter $\kappa.$ It clearly appears that the spline interpolation corresponds to the Mat\'ern prediction with $\kappa = 0.$ This limit represents an infinite range, which characterizes the smoothness of the phenomenon.

\begin{figure}[h]
\centering\includegraphics[width=1\textwidth]{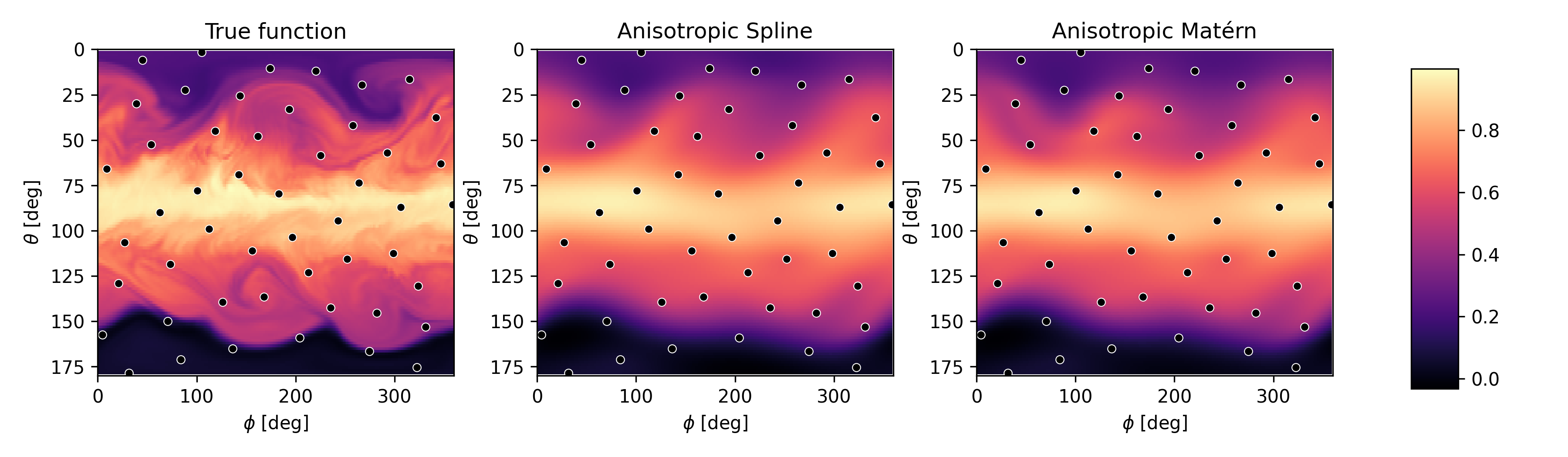}
\caption{Results for the normalized concentration of carbon dioxide on the sphere in the first scenario illustrated in 2D using the spherical coordinates $(\theta,\phi).$ $n=50$ observation points are shown as black dots. 
Left: true function. 
Middle: prediction with anisotropic finite-element splines.
Right: prediction with anisotropic mat\'ern SPDE. 
}
\label{results_matern}
\end{figure}

\begin{figure}[h]
\centering
\includegraphics[width=0.7\textwidth]{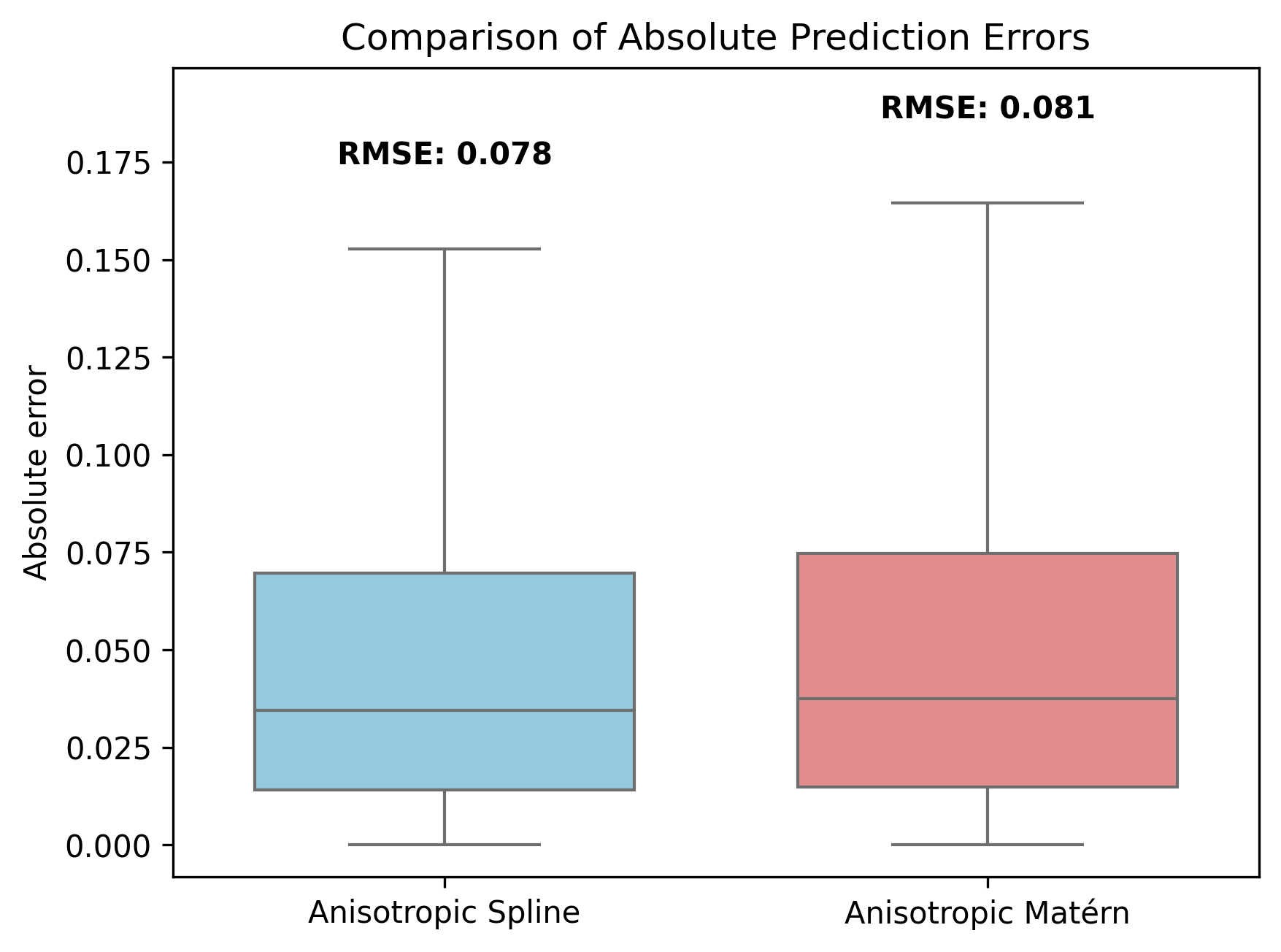}
\caption{Boxplots of absolute prediction errors at the triangulation nodes for the real-world data on the sphere, for comparison with mat\'ern prediction.}
\label{boxplots_matern}
\end{figure}

Similarly to the spline interpolation approach, a field of anisotropy parameters is considered through the introduction of pilot points (see Section~\ref{real_world}). The results are presented in Figure~\ref{results_matern}, and the corresponding errors in Figure~\ref{boxplots_matern}. They are very close to those obtained using spline interpolation, as a very small coefficient $\kappa$ can reproduce the same behavior. However, the main difficulty with Mat\'ern prediction is that the very small values of $\kappa$ required for highly smooth phenomena lead to numerical instabilities, whereas the spline prediction corresponds to the stable limit where $\kappa = 0.$

\FloatBarrier

\subsection{Analytical function on the cylinder}

\begin{figure}[h]
\centering\includegraphics[width=1\textwidth]{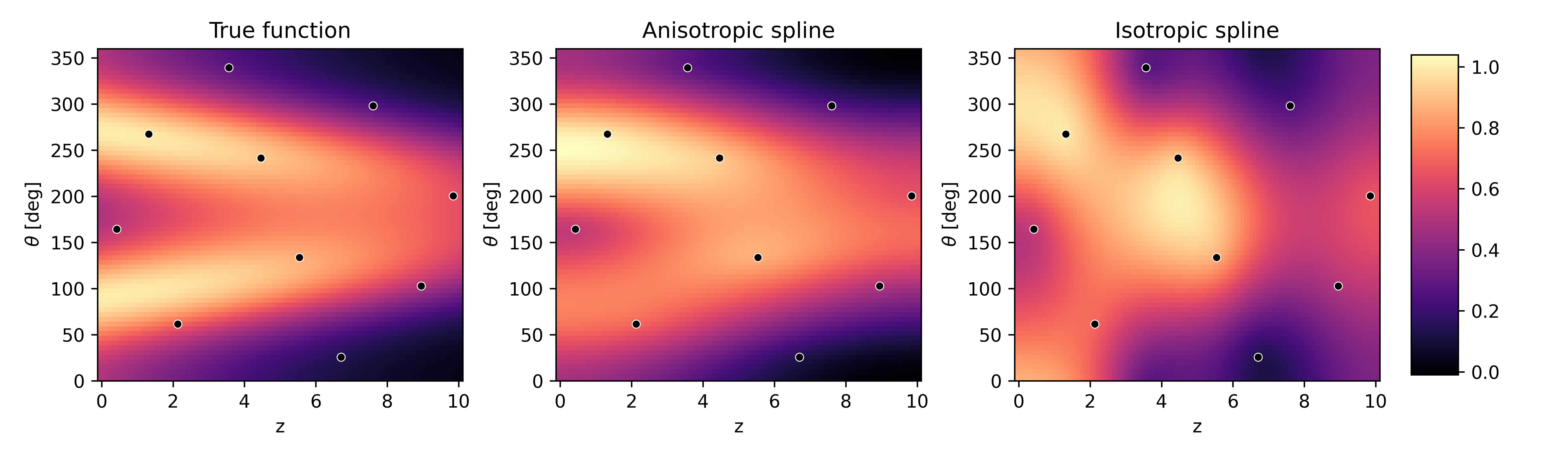}
\caption{Results for the analytical function on the cylinder in the first scenario, illustrated in 2D using the cylindrical coordinates $(\theta,z).$ $n=10$ observation points are shown as black dots. 
Left: true function. 
Middle: prediction with anisotropies induced in the local charts defined by $(\theta,z).$ 
Right: isotropic splines (no space deformation).}
\label{results_cylinder_2D}
\end{figure}

\FloatBarrier
\end{appendices}

\end{document}